\documentclass[letterpaper,12pt]{article}

\usepackage{authblk}
\usepackage[top=0.75in, bottom=1in, left=0.625in, right=0.625in]{geometry}
\usepackage{amsmath}
\usepackage{amssymb}
\usepackage{mathtools}
\usepackage{bm}
\usepackage{graphicx}
\usepackage[font={small}]{caption}
\usepackage{subcaption}
\usepackage{pifont}
\newcommand{\cmark}{\ding{51}}%
\newcommand{\xmark}{\ding{55}}%
\usepackage{tikz}
\usetikzlibrary{positioning,calc,decorations.pathreplacing,patterns,decorations.pathmorphing,decorations.markings}
\usepackage[percent]{overpic}
\usepackage{siunitx}
\usepackage{booktabs}
\usepackage[hidelinks]{hyperref}

\DeclareMathAlphabet{\mbf}{OT1}{ptm}{b}{n}
\newcommand{\mc}[1]{\ensuremath{\mathcal{#1}}}
\newcommand{\mbs}[1]{\ensuremath{\boldsymbol{#1}}}
\newcommand{\trans}{{\ensuremath{\mathsf{T}}}}
\newcommand{\frob}{{\ensuremath{\mathsf{F}}}}
\newcommand{\diag}{{\ensuremath{\mathrm{diag}}}}
\newcommand{\trace}{{\ensuremath{\mathrm{tr}}}}
\newcommand{\He}[1]{{\ensuremath{\mathrm{He}\!\left\{#1\right\}}}}

\title{Closed-loop Koopman operator approximation}
\author[1]{Steven Dahdah\thanks{E-mail: \href{mailto:steven.dahdah@mail.mcgill.ca}{steven.dahdah@mail.mcgill.ca}}}
\author[1]{James Richard Forbes}
\affil[1]{Department of Mechanical Engineering, McGill University, Montreal QC H3A~0C3, Canada}

\begin{document}

\tikzstyle{overview} = [
draw=gray,
rounded corners=0.1cm,
line width=1.5pt,
minimum height=10.5cm,
inner sep=0.1cm,
anchor=north west,
align=center,
]

\tikzstyle{block} = [
draw,
minimum width=2cm,
minimum height=1.2cm
]

\tikzstyle{smallblock} = [
draw,
minimum width=1.2cm,
minimum height=1.2cm
]

\tikzstyle{sum} = [
draw,
circle,
minimum size=0.6cm,
]

\tikzstyle{diff} = [
sum,
label=170:{\small $+$},
label=260:{\small $-$}
]

\tikzstyle{summ} = [
sum,
label=170:{\small $+$},
label=80:{\small $+$}
]

\newsavebox{\plant}
\begin{lrbox}{\plant}
    \tikzstyle{block} = [
    draw,
    minimum width=2cm,
    minimum height=1.2cm
    ]
    \begin{tikzpicture}
        \node[
        ] (in) at (0, 0) {};
        \node[
            block,
            right = 3ex of in
        ] (G) {%
            $\mbs{\mc{P}}
            \stackrel{\min}{\sim}
            \left[\begin{array}{c|c}
                \mbf{A}^{\!\mathrm{p}} & \mbf{B}^{\mathrm{p}} \\
                \hline 
                \mbf{C}^{\mathrm{p}} & \mbf{D}^{\mathrm{p}}
            \end{array}\right]$
            };
        \draw[-stealth] (in) -- (G);
        \node[
            right = 3ex of G
        ] (out) {};
        \draw[-stealth] (G) -- (out);
    \end{tikzpicture}
\end{lrbox}

\newsavebox{\closedloop}
\begin{lrbox}{\closedloop}
    \tikzstyle{block} = [
    draw,
    minimum width=2cm,
    minimum height=1.2cm
    ]
    \tikzstyle{sum} = [
    draw,
    circle,
    minimum size=0.6cm,
    ]
    \tikzstyle{diff} = [
    sum,
    label=170:{\small $+$},
    label=260:{\small $-$}
    ]
    \begin{tikzpicture}
        \node[
            block
        ] (plant) at (0, 0) {%
            $\mbs{\mc{P}}
            \stackrel{\min}{\sim}
            \left[\begin{array}{c|c}
                {\mbf{A}^{\!\mathrm{p}}} & {\mbf{B}^{\mathrm{p}}} \\
                \hline 
                \mbf{C}^{\mathrm{p}} & \mbf{D}^{\mathrm{p}}
            \end{array}\right]$
        };
        \node[
            block,
            left=3ex of plant,
        ] (cont_i) {%
            $\mbs{\mc{C}}
            \stackrel{\min}{\sim}
            \left[\begin{array}{c|c}
                \mbf{A}^{\!\mathrm{c}} & \mbf{B}^{\mathrm{c}} \\
                \hline 
                \mbf{C}^{\mathrm{c}} & \mbf{D}^{\mathrm{c}}
            \end{array}\right]$
        };
        \draw[-stealth] (cont_i) -- (plant)
            node[midway,above] {};
        \node[
            diff,
            left = 3ex of cont_i
        ] (diff_i) {};
        \draw[-stealth] (diff_i) -- (cont_i)
            node[near start,above] {};
        \draw[-stealth] (plant.east) -| ++(3ex, -8ex) -| (diff_i)
            node[near end, right] {};
        \node[
            left = 3ex of diff_i
        ] (i_bar) {$\mbf{r}_k$};
        \draw[-stealth] (i_bar) -- (diff_i);
        \node[
            right = 6ex of plant,
        ] (out) {};
        \draw[-stealth] (plant) -- (out);
        \node[
            below right = 12ex and 0ex of cont_i.south,
        ] (pic2) {%
            \begin{overpic}[width=2.5cm]{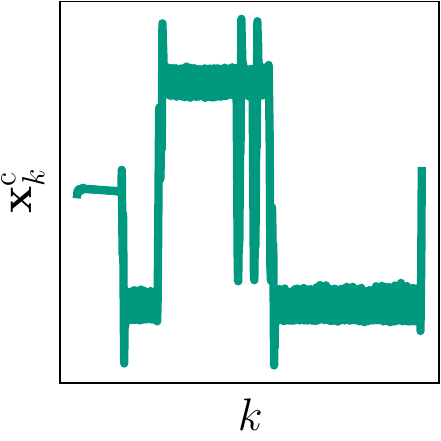}
               \put(-10, -10){\includegraphics[width=2.5cm]{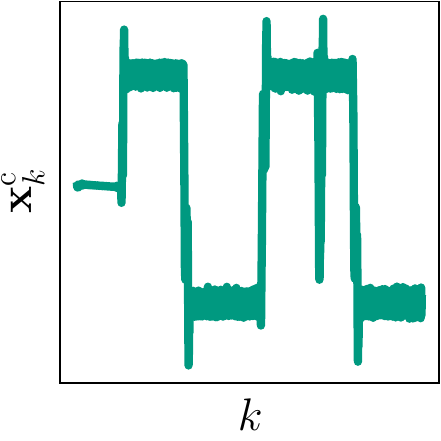}}
               \put(-20, -20){\includegraphics[width=2.5cm]{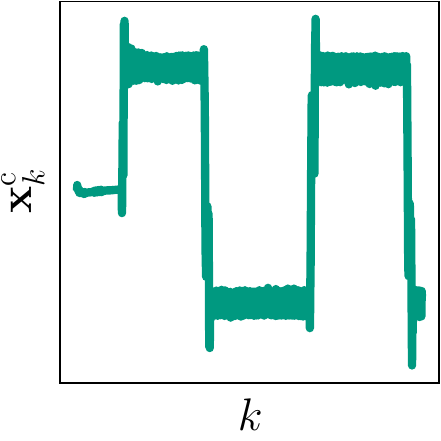}}
            \end{overpic}
        };
        \node[
            below right = 12ex and 0ex of plant.south,
        ] (pic3) {%
            \begin{overpic}[width=2.5cm]{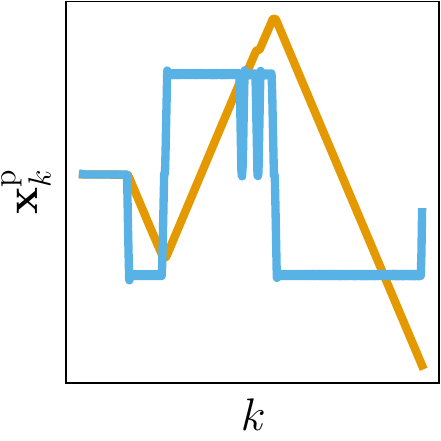}
               \put(-10, -10){\includegraphics[width=2.5cm]{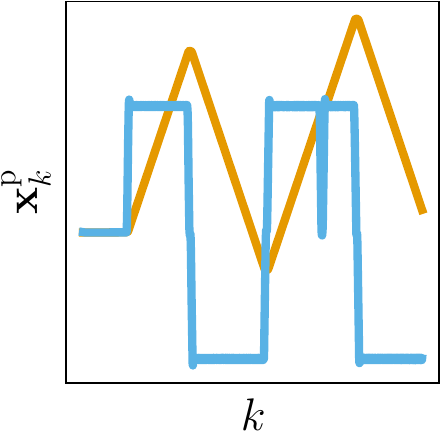}}
               \put(-20, -20){\includegraphics[width=2.5cm]{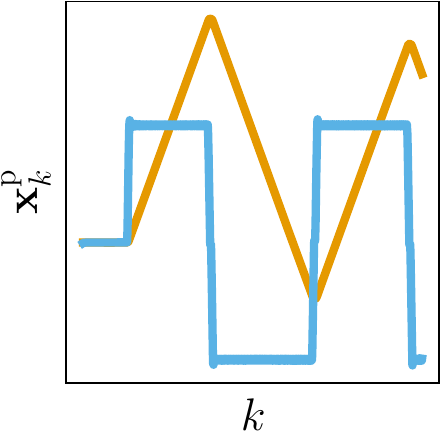}}
            \end{overpic}
        };
        \node[
            below right = 13.6ex and 0ex of i_bar.south,
        ] (pic1) {%
            \begin{overpic}[width=2.5cm]{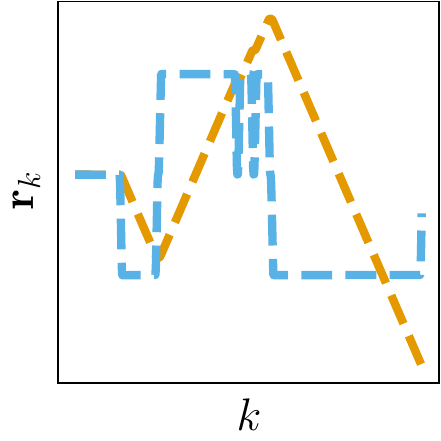}
               \put(-10, -10){\includegraphics[width=2.5cm]{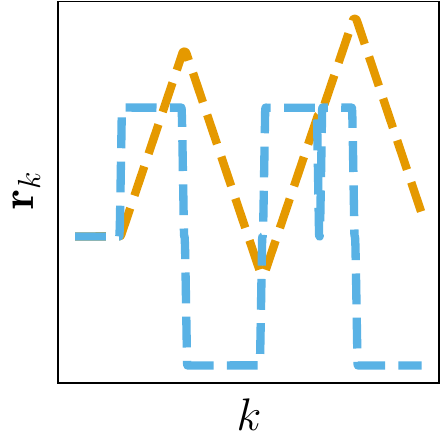}}
               \put(-20, -20){\includegraphics[width=2.5cm]{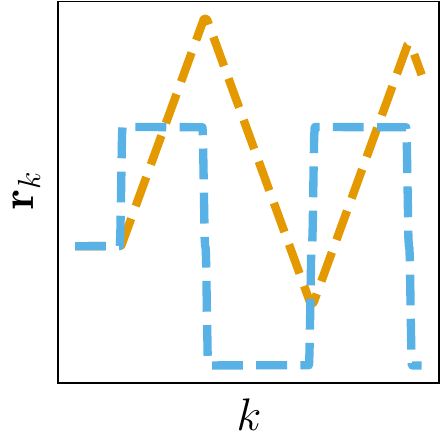}}
            \end{overpic}
        };
        \node[
            above = 0ex of pic1,
        ] (lab1) {
            \textit{Reference} $\mbf{r}_k$
        };
        \node[
            above = 0ex of pic2,
        ] (lab2) {
            \textit{Controller state} $\mbf{x}_k^{\mathrm{c}}$
        };
        \node[
            above = 0ex of pic3,
        ] (lab3) {
            \textit{Plant state $\mbf{x}_k^{\mathrm{p}}$}
        };
        \draw[
            -latex,
            line width=2pt,
            shorten <= 0ex,
            shorten >= 0ex,
        ] (i_bar) -- (lab1);
        \draw[
            -latex,
            line width=2pt,
            shorten <= 1ex,
            shorten >= 0ex,
        ] (cont_i) -- (lab2);
        \draw[
            -latex,
            line width=2pt,
            shorten <= 1ex,
            shorten >= 0ex,
        ] (plant) -- (lab3);
    \end{tikzpicture}
\end{lrbox}

\newsavebox{\arrow}
\begin{lrbox}{\arrow}
    \begin{tikzpicture}
        \draw[-latex, line width=2pt] (0cm, 0cm) -- (0cm, -1.0cm);
    \end{tikzpicture}
\end{lrbox}

\maketitle

\begin{abstract}
    This paper proposes a method to identify a Koopman model of a
    feedback-controlled system given a known controller.
    The Koopman operator allows a nonlinear system to be rewritten
    as an infinite-dimensional linear system by viewing it in terms of an
    infinite set of lifting functions.
    A finite-dimensional approximation of the Koopman operator can be identified
    from data by choosing a finite subset of lifting functions and solving a
    regression problem in the lifted space.
    Existing methods are designed to identify open-loop systems. However, it is
    impractical or impossible to run experiments on some systems, such as
    unstable systems, in an open-loop fashion.
    The proposed method leverages the linearity of the Koopman operator, along
    with knowledge of the controller and the structure of the closed-loop
    system, to simultaneously identify the closed-loop and plant systems.
    The advantages of the proposed closed-loop Koopman operator approximation
    method are demonstrated in simulation using a Duffing oscillator
    and experimentally using a rotary inverted pendulum system.
    An open-source software implementation of the proposed method is publicly
    available, along with the experimental dataset generated for this paper.
\end{abstract}


\section{Introduction}\label{sec:introduction}

Using Koopman operator theory~\cite{koopman_hamiltonian_1931,
    mezic_2019_spectrum, budisic_applied_2012, mauroy_2020_koopman}, a
finite-dimensional nonlinear system can be rewritten as an infinite-dimensional
linear system.
The Koopman operator itself serves to advance the infinite-dimensional system's
state in time, much like the dynamics matrix of a linear system.
Practically working with an infinite-dimensional operator is intractable, so the
Koopman operator is approximated as a finite-dimensional matrix, often using
data-driven methods.
While the Koopman operator was originally proposed almost one hundred years ago,
recent theoretical results~\cite{mezic_2019_spectrum, budisic_applied_2012,
    mauroy_2020_koopman} paired with modern computational resources have led to it
becoming a popular tool for identifying dynamical systems from data.

In a typical Koopman operator approximation workflow, a set of lifting functions
is first chosen. In some situations, these lifting functions are inspired by the
dynamics of the system~\cite{abraham_active_2019, mamakoukas_local_2019,
    kaiser_data_2021}, while in others, they are taken from a standard set of basis
functions, like polynomials, sinusoids, or radial basis
functions~\cite{bruder_modeling_2019, abraham_model-based_2017,
    mallen_koopman_2023}. Lifting functions can also be chosen to approximate a
given kernel~\cite{guo_2022_koopman, degennaro_2019_scalable,
    rahimi_2007_random}. Time delays are often incorporated into the lifted state as
well~\cite{korda_2018_linear, bruder_modeling_2019, pan_2020_structure}.
However, finding a practical set of Koopman lifting functions for a
given system is often a significant challenge, both for theoretical and
numerical reasons. The existence of useful sets of Koopman lifting functions is
analyzed theoretically in~\cite{arathoon2, liu1}.
Once the chosen lifting functions are applied to the data, linear regression is
used to approximate the Koopman matrix, often with some form of
regularization~\cite{bruder_modeling_2019}.

One of the key properties of the Koopman operator is its linearity, which makes
it possible to leverage existing linear systems theory when identifying or
analyzing Koopman systems.
The linear Koopman representation of the nonlinear system also allows linear
control design methods to be leveraged~\cite{korda_2018_linear,
    otto_2021_koopman, abraham_active_2019, mamakoukas_local_2019,
    bruder_modeling_2019, uchida_2021_data-driven, kaiser_data_2021}. Koopman
operator approximation methods are also closely related to classical system
identification methods~\cite{korda_2018_linear, ljung_2019_a}, most notably
subspace identification methods~\cite[\S3]{brunton_2022_modern}.
This similarity means that well-established system identification techniques can
often be adapted to enrich the Koopman workflow.

An area of particular interest is the identification of closed-loop (CL)
systems~\cite{forssell_1999_closedloop, hof_1998_closedloop,
    overschee_1997_closed, ljung_1999_system}.
It is often unsafe or impractical to run an experiment on a plant without a
feedback control system~\cite[\S1.1, \S17.3]{ljung_1999_system}.
In some cases, the control loop is simply ignored and the controller's outputs
are treated as exogenous input signals to the plant. However, this
simplification, known as the \textit{direct approach} to closed-loop system
identification~\cite[\S13.5]{ljung_1999_system}, is problematic in some
situations.
The feedback loop introduces correlations between the plant's output and its
input, leading to biased estimates of the plant's
dynamics~\cite{forssell_1999_closedloop, hof_1998_closedloop},~\cite[\S11.1]{katayama_2005_subspace}.
In cases where this bias is small, the direct approach may yield acceptable
results. In cases where the bias is not small, more sophisticated closed-loop
system identification methods must be employed.
The \textit{indirect approach} to closed-loop system
identification~\cite[\S13.5]{ljung_1999_system}, which uses the reference signal
as the exogenous input, identifies the closed-loop system as a whole. The
indirect approach then uses knowledge of the controller to extract a model of
the plant system for use on its
own~\cite{forssell_1999_closedloop, hof_1998_closedloop},~\cite[\S11.1]{katayama_2005_subspace}.
While closed-loop system identification can lead to more complex plant models,
in many situations it is the appropriate tool to use to avoid biased estimates
of model parameters.

Closed-loop identification is particularly practical when identifying unstable
plants. Evaluating the performance of an unstable model is challenging, as
prediction errors may diverge even if the identified model is accurate.
This is particularly problematic in the Koopman framework, where lifting
functions are unknown and are often chosen based on the prediction error they
achieve.
By first identifying a model of the closed-loop system, which is assumed to be
asymptotically stable, the closed-loop prediction error can be used as a more
informative goodness-of-fit metric~\cite[\S4.1]{forssell_1999_closedloop}.
Closed-loop identification methods also allow the closed-loop system's model
parameters to be regularized, when regularizing the unstable plant's parameters
may incorrectly lead to a stable model.

\begin{figure}[!t]
    \centering
    \resizebox{\textwidth}{!}{%
        \begin{tikzpicture}
            \node[
                overview,
                minimum width=4.5cm,
                label={north:(a) Collect data snapshots from closed-loop system},
            ] (step1) at (0, 0) {%
                \usebox{\closedloop}\\[5ex]
                \textit{Known:}
                $\mbs{\mc{C}}
                \stackrel{\min}{\sim}
                \left[\begin{array}{c|c}
                    \mbf{A}^{\!\mathrm{c}} & \mbf{B}^{\mathrm{c}} \\
                    \hline 
                    \mbf{C}^{\mathrm{c}} & \mbf{D}^{\mathrm{c}}
                \end{array}\right],\ 
                \mbf{C^\mathrm{p}} = \diag\left\{\mbf{1}, \mbf{0}\right\},\ 
                \mbf{D^\mathrm{p}} = \mbf{0}$ \\[2ex]
                \textit{Unknown:}
                $\mbf{U}^\mathrm{p}
                =
                \begin{bmatrix}
                    \mbf{A^{\!\mathrm{p}}} & \mbf{B^\mathrm{p}}
                \end{bmatrix}$
            };
            \node[
                overview,
                right=0.25cm of step1,
                label={north:(b) Lift and arrange snapshots},
            ] (step2) {%
                \textit{Lifted plant state:}\\[1ex]
                \begin{overpic}[width=3cm]{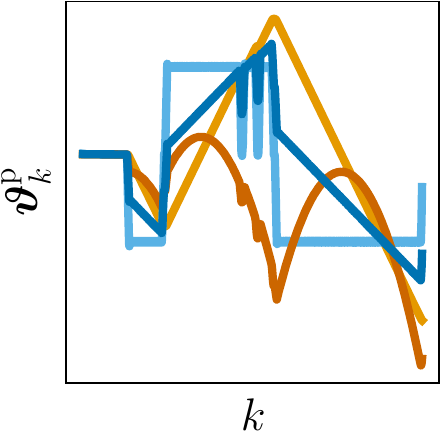}
                   \put(-10, -10){\includegraphics[width=3cm]{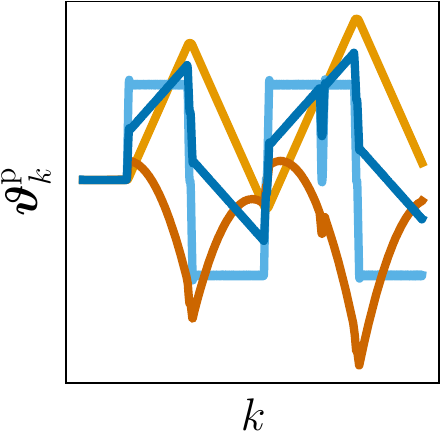}}
                   \put(-20, -20){\includegraphics[width=3cm]{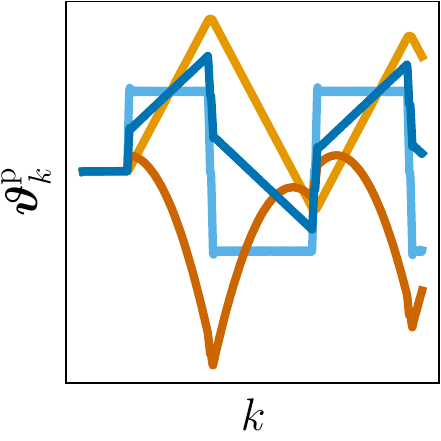}}
                \end{overpic}\\[3ex]
                \usebox{\arrow}
                \\
                \textit{Arranged lifted snapshots:}\\[2ex]
                $\mbs{\Psi} = \begin{bmatrix}
                    \mbf{x}_0^\mathrm{c} &
                        \mbf{x}_1^\mathrm{c} &
                        \cdots &
                        \mbf{x}_{q-1}^\mathrm{c} \\[1ex]
                    \mbs{\vartheta}_0^\mathrm{p} &
                        \mbs{\vartheta}_1^\mathrm{p} &
                        \cdots &
                        \mbs{\vartheta}_{q-1}^\mathrm{p} \\
                    \mbf{r}_0 &
                        \mbf{r}_1 &
                        \cdots &
                        \mbf{r}_{q-1}
                \end{bmatrix}$ \\[2ex]
                $\mbs{\Theta}_+ = \begin{bmatrix}
                    \mbf{x}_1^\mathrm{c} &
                        \mbf{x}_2^\mathrm{c} &
                        \cdots &
                        \mbf{x}_{q}^\mathrm{c} \\[1ex]
                    \mbs{\vartheta}_1^\mathrm{p} &
                        \mbs{\vartheta}_2^\mathrm{p} &
                        \cdots &
                        \mbs{\vartheta}_{q}^\mathrm{p}
                \end{bmatrix}$
            };
            \node[
                overview,
                right=0.25cm of step2,
                label={north:(c) Identify CL and plant systems},
            ] (step4) {%
                \textit{CL Koopman matrix structure:}\\[1ex]
                $\mbf{U}^{\mathrm{f}}
                =
                \begin{bmatrix}
                    \mbf{A}^{\!\mathrm{c}} &
                    - \mbf{B}^{\mathrm{c}} \mbf{C}^{\mathrm{p}} &
                    \mbf{B}^{\mathrm{c}}
                    \\
                    \mbf{B}^{\mathrm{p}} \mbf{C}^{\mathrm{c}} &
                    \mbf{A}^{\!\mathrm{p}}
                    - \mbf{B}^{\mathrm{p}} \mbf{D}^{\mathrm{c}} \mbf{C}^{\mathrm{p}} &
                    \mbf{B}^{\mathrm{p}} \mbf{D}^{\mathrm{c}}
                \end{bmatrix}$
                \\[1ex]
                \usebox{\arrow}
                \\
                \textit{Constrained EDMD:}\\[1ex]
                $\begin{aligned}
                    \min\;&
                    J(\mbf{U}^{\mathrm{f}}, \mbf{U}^{\mathrm{p}})
                    =
                    \|\mbs{\Theta}_+ - \mbf{U}^{\mathrm{f}}\mbs{\Psi}\|_\frob^2
                    \\
                    \mathrm{s.t.}\;&
                    \begin{bmatrix}
                        \mbf{U}_{21}^{\mathrm{f}} &
                        \mbf{U}_{23}^{\mathrm{f}}
                    \end{bmatrix}
                    =
                    \mbf{B}^{\mathrm{p}}
                    \begin{bmatrix}
                        \mbf{C}^{\mathrm{c}} &
                        \mbf{D}^{\mathrm{c}}
                    \end{bmatrix},
                    \\
                    &\mbf{U}_{22}^{\mathrm{f}}
                    =
                    \mbf{A}^{\!\mathrm{p}}
                    -
                    \mbf{B}^{\mathrm{p}}
                    \mbf{D}^{\mathrm{c}}
                    \mbf{C}^{\mathrm{p}}
                \end{aligned}$ \\[1ex]
                \usebox{\arrow}
                \\
                \textit{Plant Koopman system:}\\[1ex]
                \usebox{\plant}
            };
        \end{tikzpicture}
    }
    \caption{Overview of the proposed closed-loop Koopman operator
    identification method. To simplify the presentation, no feedforward signal
    is used.\@ (a) First, the controller reference, controller state, and plant
    state are recorded during a series of experiments. The controller state
    space matrices are known, and $\mbf{C}^\mathrm{p}$ and $\mbf{D}^\mathrm{p}$
    are fixed. If the controller state is not directly available, it can be
    computed from its input and output. Only the Koopman matrix of the plant
    $\mbf{U}^\mathrm{p}$ is unknown.\@ (b) The plant state is lifted and
    augmented with the controller state and reference.\@ (c) The Koopman
    matrices of the closed-loop (CL) system and the plant system are
    approximated simultaneously by incorporating the known structure of the
    closed-loop system as a constraint on the Extended DMD (EDMD)
    problem.}\label{fig:overview}
\end{figure}
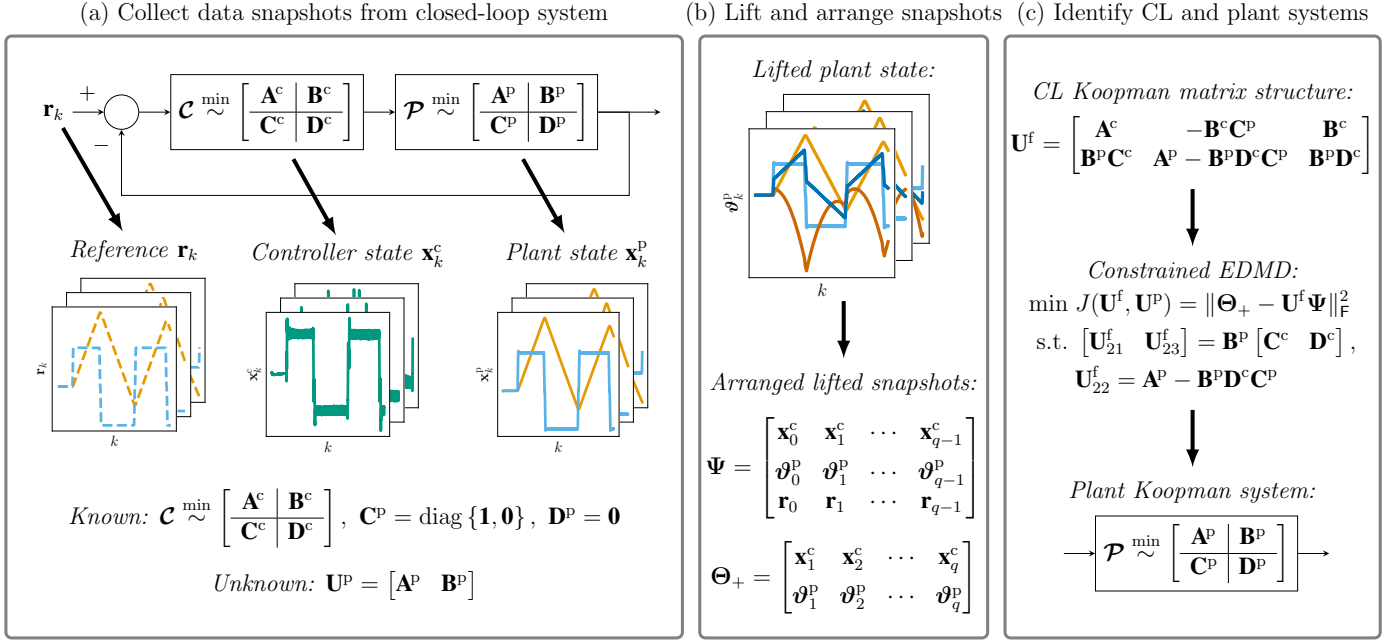

The key contribution of this paper is a means to identify a Koopman
representation of a system using closed-loop data. The proposed method,
summarized in Figure~\ref{fig:overview}, is a variation of the indirect approach
to system identification, wherein Koopman models of the closed-loop system and
the plant system are simultaneously identified given knowledge of the
controller.
The significance of this work is that a system that cannot be operated in
open-loop without a feedback controller can now be identified using Koopman
operator theory.
Even in situations where open-loop or closed-loop methods would identify the
same model, it is shown that the closed-loop system is more convenient to work
with, as it naturally provides a bounded prediction error and allows the
closed-loop Koopman matrix to be included in a regularizer or additional
constraint, such as an asymptotic stability
constraint~\cite{dahdah_system_2022}.
The advantages of the proposed method are demonstrated on a simulated
closed-loop Duffing oscillator system subject to coloured measurement noise,
as well as on a rotary inverted pendulum system which, due to its instability,
would be difficult to identify in an open-loop setting.

The remainder of this paper is as follows.
Section~\ref{sec:background} outlines the necessary background information on
the Koopman operator and its associated approximation techniques.
Section~\ref{sec:closed_loop} derives the proposed closed-loop Koopman operator
approximation method.
Sections~\ref{sec:duffing_example} and~\ref{sec:results} analyze the
simulated and experimental results of the closed-loop Koopman operator
approximation method respectively.
Finally, Section~\ref{sec:conclusion} presents the paper's conclusions and
discusses avenues for future research.

\section{Background}\label{sec:background}
\subsection{Koopman operator theory}
Consider the nonlinear difference equation
\begin{equation}
    \mbf{x}_{k+1} = \mbf{f}(\mbf{x}_{k}),%
    \label{eq:dynamics_no_input}
\end{equation}
where
${\mbf{x}_{k} \in \mc{M}}$
evolves on a manifold
${\mc{M} \subseteq \mathbb{R}^{m \times 1}}$.
Also consider an infinite number of scalar-valued \textit{lifting functions},
${\psi: \mc{M} \to \mathbb{R}}$,
which span an infinite-dimensional Hilbert space
$\mc{H}$.
The \textit{Koopman operator}, ${\mc{U}: \mc{H} \to \mc{H}}$, is a linear
operator that composes all lifting functions ${\psi \in \mc{H}}$ with
$\mbf{f}(\cdot)$, thereby advancing them in time by
one step. That is~\cite[\S3.2]{kutz_dynamic_2016},
\begin{equation}
    (\mc{U} \psi)(\cdot) = (\psi \circ \mbf{f})(\cdot).%
    \label{eq:koopman_def_no_input}
\end{equation}
Evaluating~\eqref{eq:koopman_def_no_input} at $\mbf{x}_k$ reveals that the
dynamics of~\eqref{eq:dynamics_no_input} can be rewritten linearly in terms of
$\psi$ as
\begin{equation}
    \psi(\mbf{x}_{k+1}) = (\mc{U} \psi)(\mbf{x}_{k}).%
    \label{eq:koopman_dynamics}
\end{equation}
A finite-dimensional approximation of~\eqref{eq:koopman_dynamics} is
\begin{equation}
    \mbs{\psi}(\mbf{x}_{k+1}) = \mbf{U} \mbs{\psi}(\mbf{x}_{k}) + \mbs{\epsilon}_k,%
    \label{eq:koopman_approx_no_input}
\end{equation}
where
${\mbs{\psi}: \mc{M} \to \mathbb{R}^{p \times 1}}$ is the \textit{vector-valued
    lifting function}, ${\mbf{U} \in \mathbb{R}^{p \times p}}$ is the
\textit{Koopman matrix}, and $\mbs{\epsilon}_k$ is the residual error.

\subsection{Koopman operator theory with inputs}
The definition of the Koopman operator can be modified to accommodate nonlinear
difference equations with exogenous inputs. Consider the difference equation
\begin{equation}
    \mbf{x}_{k+1} = \mbf{f}(\mbf{x}_{k}, \mbf{u}_k),\label{eq:dynamics_input}
\end{equation}
where the state is
${\mbf{x}_{k} \in \mc{M} \subseteq \mathbb{R}^{m \times 1}}$
and the input is
${\mbf{u}_{k} \in \mc{N} \subseteq \mathbb{R}^{n \times 1}}$.
The lifting functions are now
${\psi: \mc{M} \times \mc{N} \to \mathbb{R}}$
and the Koopman operator
${\mc{U}: \mc{H} \to \mc{H}}$
now satisfies
\begin{equation}
    (\mc{U} \psi)(\mbf{x}_{k}, \mbf{u}_{k})
    = \psi(\mbf{f}(\mbf{x}_{k}, \mbf{u}_k), \star),
    \label{eq:koopman_def_input}
\end{equation}
where
${\star = \mbf{u}_k}$
if the input has state-dependent dynamics, or
${\star = \mbf{0}}$
if the input has no dynamics~\cite[\S6.5]{kutz_dynamic_2016}.
Let the vector-valued lifting function
${\mbs{\psi}: \mc{M} \times \mc{N} \to \mathbb{R}^{p \times 1}}$
be partitioned as
\begin{equation}
    \mbs{\psi}(\mbf{x}_k, \mbf{u}_k) = \begin{bmatrix}
        \mbs{\vartheta}(\mbf{x}_k) \\
        \mbs{\upsilon}(\mbf{x}_k, \mbf{u}_k)
    \end{bmatrix},
\end{equation}
where the state-dependent lifting functions are
${\mbs{\vartheta}: \mc{M} \to \mathbb{R}^{p_\vartheta \times 1}}$,
the input-dependent lifting functions are
${\mbs{\upsilon}: \mc{M} \times \mc{N} \to \mathbb{R}^{p_\upsilon \times 1}}$,
and
the lifting function dimensions satisfy
${p_\vartheta + p_\upsilon = p}$.
With an exogenous input, substituting the state and input
into~\eqref{eq:koopman_def_input} results in~\cite[\S6.5.1]{kutz_dynamic_2016}
\begin{equation}
    \mbs{\vartheta}(\mbf{x}_{k+1}) \\
    =
    \mbf{U}
    \mbs{\psi}(\mbf{x}_{k}, \mbf{u}_{k})
    + \mbs{\epsilon}_k,%
    \label{eq:U_part}
\end{equation}
where ${\mbf{U} = \begin{bmatrix} \mbf{A} & \mbf{B} \end{bmatrix}}$.
Expanding~\eqref{eq:U_part} yields the familiar linear state-space form,
\begin{equation}
    \mbs{\vartheta}(\mbf{x}_{k+1})
    =
    \mbf{A} \mbs{\vartheta}(\mbf{x}_{k})
    + \mbf{B} \mbs{\upsilon}(\mbf{x}_k, \mbf{u}_k)
    + \mbs{\epsilon}_k.
\end{equation}

When designing lifting functions, the first $m$ lifted states are often chosen
to be the state of the original difference equation,
$\mbf{x}_k$. Specifically,~\cite[\S3.3.1]{kutz_dynamic_2016}
\begin{equation}
    \mbs{\vartheta}(\mbf{x}_k)
    =
    \begin{bmatrix}
        \mbf{x}_k \\
        \vartheta_m(\mbf{x}_k) \\
        \vartheta_{m + 1}(\mbf{x}_k) \\
        \vdots \\
        \vartheta_{p_\vartheta - 1}(\mbf{x}_k)
    \end{bmatrix}.
\end{equation}
This choice of lifting functions makes it easier to recover the original state
from the lifted state.
Another common design decision is to leave the input unlifted when identifying
a Koopman model for control. That is~\cite{korda_2018_linear},
\begin{equation}
    \mbs{\upsilon}(\mbf{x}_k, \mbf{u}_k) = \mbf{u}_k.
\end{equation}
However, recent work has also shown that bilinear input-dependent lifting
functions are a better alternative for control affine
systems~\cite{bruder_2021_advantages}.

An output equation,
\begin{equation}
    \mbs{\zeta}_k = \mbf{C} \mbs{\vartheta}_k + \mbf{D} \mbs{\upsilon}_k,
\end{equation}
where $\mbs{\zeta}_k \in \mathbb{R}^{p_\zeta \times 1}$,
can also be considered to incorporate the Koopman operator into a true linear
system with input, output, and state. While $\mbf{D} = \mbf{0}$ in all cases,
$\mbf{C}$ can be chosen to recover the original states, or any other desired
output.
If the original state is not directly included in the lifted state, $\mbf{C}$
can instead be determined using least-squares~\cite[\S3.2.1]{korda_2018_linear}.
The \textit{Koopman system} is therefore
\begin{equation}
    \mbs{\mc{G}}
    \stackrel{\min}{\sim}
    \left[\begin{array}{c|c} 
        \mbf{A} & \mbf{B} \\
        \hline 
        \mbf{C} & \mbf{D}
    \end{array}\right],
\end{equation}
where $\stackrel{\min}{\sim}$ denotes a minimal state space
realization~\cite[\S3.2.1]{green_linear_1994}.

\subsection{Approximation of the Koopman operator from data}
To approximate the Koopman matrix from a dataset
${\mc{D} = {\{\mbf{x}_k, \mbf{u}_k\}}_{k=0}^q}$,
consider the lifted snapshot matrices
\begin{align}
    \mbs{\Psi} &= \begin{bmatrix}
        \mbs{\psi}_{0} & \mbs{\psi}_{1} & \cdots & \mbs{\psi}_{q-1}
    \end{bmatrix} \in \mathbb{R}^{p \times q}, \\
    \mbs{\Theta}_+ &= \begin{bmatrix}
        \mbs{\vartheta}_{1} & \mbs{\vartheta}_{2} & \cdots & \mbs{\vartheta}_{q}
    \end{bmatrix} \in \mathbb{R}^{p_\vartheta \times q},\label{eq:Theta}
\end{align}
where
${\mbs{\psi}_k = \mbs{\psi}(\mbf{x}_k, \mbf{u}_k)}$
and
${\mbs{\vartheta}_k = \mbs{\vartheta}(\mbf{x}_k)}$.
The Koopman matrix that minimizes
\begin{equation}
    J(\mbf{U}) = \frac{1}{q} \|\mbs{\Theta}_+ - \mbf{U} \mbs{\Psi}\|_\frob^2
\end{equation}
is~\cite[\S1.2.1]{kutz_dynamic_2016}
\begin{equation}
    \mbf{U} = \mbs{\Theta}_+ \mbs{\Psi}^\dagger,\label{eq:pseudoinverse}
\end{equation}
where ${(\cdot)}^\dagger$ denotes the Moore-Penrose pseudoinverse.

When the dataset contains many snapshots, the pseudoinverse required
in~\eqref{eq:pseudoinverse} is numerically ill-conditioned. Extended Dynamic
Mode Decomposition (EDMD)~\cite{williams_data-driven_2015} can alleviate this
problem when dataset contains many fewer states than snapshots
(\textit{i.e.}, when $p \ll q$)~\cite[\S10.3]{kutz_dynamic_2016}.
The EDMD approximation of the Koopman matrix is
\begin{equation}
    \mbf{U}
    =
    \mbs{\Theta}_+
    \left(
        \mbs{\Psi}^\trans
        \mbs{\Psi}^{\trans^\dagger}
    \right)
    \mbs{\Psi}^\dagger
    =
    \left(\mbs{\Theta}_+ \mbs{\Psi}^\trans\right)
    {\left(\mbs{\Psi} \mbs{\Psi}^\trans\right)}^\dagger
    =
    \mbf{G}
    \mbf{H}^\dagger,
    \label{eq:edmdi-sol}
\end{equation}
where
\begin{equation}
    \mbf{G}
    =
    \frac{1}{q}
    \mbs{\Theta}_+ \mbs{\Psi}^\trans \in \mathbb{R}^{p_\vartheta \times p},
    \quad
    \mbf{H}
    =
    \frac{1}{q}
    \mbs{\Psi} \mbs{\Psi}^\trans \in \mathbb{R}^{p \times p}.
    \label{eq:edmdi-sol-scaled}
\end{equation}

Often, Tikhonov regularization is also included to improve the numerical
conditioning of $\mbf{U}$ by penalizing its squared Frobenius
norm~\cite{tikhonov_1995_numerical}. The cost function then becomes
\begin{equation}
    J(\mbf{U}) = \frac{1}{q} \|\mbs{\Theta}_+ - \mbf{U} \mbs{\Psi}\|_\frob^2
     + \frac{\alpha}{q} \|\mbf{U}\|_\frob^2,
 \end{equation}
where $\alpha$ is the regularization coefficient. The incorporation of Tikhonov
regularization into EDMD is discussed in Section~\ref{sec:closed_loop}.

\section{Closed-loop Koopman operator approximation}\label{sec:closed_loop}

In this section, the proposed closed-loop Koopman operator approximation method
is outlined in detail. First, the Koopman representation of the closed-loop
system is derived in terms of the known state-space matrices of the controller
and the unknown Koopman matrix of the plant. Then, a method for simultaneously
identifying the closed-loop system and plant system is described.
\subsection{Formulation of the closed-loop Koopman system}
\begin{figure}[htb]
    \centering
    \begin{tikzpicture}
        \node[
        ] (in) at (0, 0) {$\mbf{u}_k^{\mathrm{c}}$};

        \node[
            above = 6ex of in
        ] (ff) at (0, 0) {$\mbf{f}_k$};

        \draw[
            decorate,
            decoration = {brace},
            line width = 1pt,
        ] ([xshift=-2ex]in.south) -- ([xshift=-2ex]ff.north)
        node[
            midway, left
        ] {$\mbs{\upsilon}_k^{\mathrm{s}}=$};

        \node[
            smallblock,
            right = 4ex of in
        ] (G) {$\mbs{\mc{C}}$};
        \draw[-stealth] (in) -- (G);

        \node[
            summ,
            right = 4ex of G,
        ] (sum) {};

        \node[
            smallblock,
            right = 4ex of sum
        ] (W) {$\mbs{\mc{P}}$};

        \node[
            right = 4ex of W
        ] (out) {$\mbs{\zeta}_k^\mathrm{p} = \mbs{\zeta}_k$};

        \draw[-stealth] (G) -- (sum)
            node[midway, below] {$\mbf{y}_k^{\mathrm{c}}$};
        \draw[-stealth] (sum) -- (W)
            node[midway, below] {$\mbs{\upsilon}_k^{\mathrm{p}}$};
        \draw[-stealth] (ff) -| (sum);
        \draw[-stealth] (W) -- (out);
    \end{tikzpicture}
    \caption{Series interconnection of the controller and
    plant.}\label{fig:series_interconnection}
\end{figure}
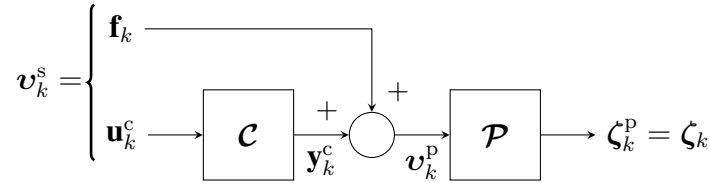

Consider the Koopman system modelling the plant,
\begin{align}
    \mbs{\vartheta}_{k+1}^{\mathrm{p}}
    &=
    \mbf{A}^{\!\mathrm{p}}
    \mbs{\vartheta}_{k}^{\mathrm{p}}
    +
    \mbf{B}^{\mathrm{p}}
    \mbs{\upsilon}_{k}^{\mathrm{p}},
    \\
    \mbs{\zeta}_{k}^{\mathrm{p}}
    &=
    \mbf{C}^{\mathrm{p}}
    \mbs{\vartheta}_{k}^{\mathrm{p}}
    +
    \mbf{D}^{\mathrm{p}}
    \mbs{\upsilon}_{k}^{\mathrm{p}},
\end{align}
along with the linear system modelling the known controller,
\begin{align}
    \mbf{x}_{k+1}^{\mathrm{c}}
    &=
    \mbf{A}^{\!\mathrm{c}}
    \mbf{x}_{k}^{\mathrm{c}}
    +
    \mbf{B}^{\mathrm{c}}
    \mbf{u}_{k}^{\mathrm{c}},
    \\
    \mbf{y}_{k}^{\mathrm{c}}
    &=
    \mbf{C}^{\mathrm{c}}
    \mbf{x}_{k}^{\mathrm{c}}
    +
    \mbf{D}^{\mathrm{c}}
    \mbf{u}_{k}^{\mathrm{c}}.
\end{align}
Let
\begin{equation}
    \mbs{\upsilon}_{k}^{\mathrm{p}}
    =
    \mbf{y}_{k}^{\mathrm{c}}
    +
    \mbf{f}_{k},
\end{equation}
which yields the series interconnection of the controller and the plant with
a feedforward signal $\mbf{f}_k$.
This feedforward signal is entirely exogenous, and could be generated by a
nonlinear inverse model of the plant.
The new Koopman system's input includes the controller input and feedforward
input, resulting in
\begin{equation}
    \mbs{\upsilon}_{k}^{\mathrm{s}}
    =
    \begin{bmatrix}
        \mbf{u}_{k}^{\mathrm{c}} \\
        \mbf{f}_{k}
    \end{bmatrix},\label{eq:upsilon_s}
\end{equation}
while its output is simply the plant output,
${\mbs{\zeta}_{k} = \mbs{\zeta}_{k}^{\mathrm{p}}}$.
The new system's state includes the controller state and plant state, resulting
in
\begin{equation}
    \mbs{\vartheta}_k
    =
    \begin{bmatrix}
        \mbf{x}_{k}^{\mathrm{c}} \\
        \mbs{\vartheta}_{k}^{\mathrm{p}}
    \end{bmatrix}.
\end{equation}
The cascaded plant and controller systems are depicted
in Figure~\ref{fig:series_interconnection}.

The state space representation of the plant becomes
\begin{align}
    \mbs{\vartheta}_{k+1}^{\mathrm{p}}
    &=
    \mbf{A}^{\!\mathrm{p}}
    \mbs{\vartheta}_{k}^{\mathrm{p}}
    +
    \mbf{B}^{\mathrm{p}}
    \left(
        \mbf{C}^{\mathrm{c}}
        \mbf{x}_{k}^{\mathrm{c}}
        +
        \mbf{D}^{\mathrm{c}}
        \mbf{u}_{k}^{\mathrm{c}}
        +
        \mbf{f}_k
    \right)
    \\
    &=
    \begin{bmatrix}
        \mbf{B}^{\mathrm{p}} \mbf{C}^{\mathrm{c}} &
        \mbf{A}^{\!\mathrm{p}}
    \end{bmatrix}
    \begin{bmatrix}
        \mbf{x}_{k}^{\mathrm{c}} \\
        \mbs{\vartheta}_{k}^{\mathrm{p}}
    \end{bmatrix}
    +
    \begin{bmatrix}
        \mbf{B}^{\mathrm{p}}
        \mbf{D}^{\mathrm{c}}
        &
        \mbf{B}^{\mathrm{p}}
    \end{bmatrix}
    \begin{bmatrix}
        \mbf{u}_{k}^{\mathrm{c}} \\
        \mbf{f}_k
    \end{bmatrix},
    \\
    \mbs{\zeta}_{k}^{\mathrm{p}}
    &=
    \mbf{C}^{\mathrm{p}}
    \mbs{\vartheta}_{k}^{\mathrm{p}}
    +
    \mbf{D}^{\mathrm{p}}
    \left(
        \mbf{C}^{\mathrm{c}}
        \mbf{x}_{k}^{\mathrm{c}}
        +
        \mbf{D}^{\mathrm{c}}
        \mbf{u}_{k}^{\mathrm{c}}
        +
        \mbf{f}_k
    \right)
    \\
    &=
    \begin{bmatrix}
        \mbf{D}^{\mathrm{p}} \mbf{C}^{\mathrm{c}} &
        \mbf{C}^{\mathrm{p}}
    \end{bmatrix}
    \begin{bmatrix}
        \mbf{x}_{k}^{\mathrm{c}} \\
        \mbs{\vartheta}_{k}^{\mathrm{p}}
    \end{bmatrix}
    +
    \begin{bmatrix}
        \mbf{D}^{\mathrm{p}}
        \mbf{D}^{\mathrm{c}}
        &
        \mbf{D}^{\mathrm{p}}
    \end{bmatrix}
    \begin{bmatrix}
        \mbf{u}_{k}^{\mathrm{c}} \\
        \mbf{f}_k
    \end{bmatrix}.
\end{align}
The state space representation of the series-interconnected system is therefore
\begin{align}
    \begin{bmatrix}
        \mbf{x}_{k+1}^{\mathrm{c}} \\
        \mbs{\vartheta}_{k+1}^{\mathrm{p}}
    \end{bmatrix}
    &=
    \begin{bmatrix}
        \mbf{A}^{\!\mathrm{c}} &
        \mbf{0} \\
        \mbf{B}^{\mathrm{p}} \mbf{C}^{\mathrm{c}} &
        \mbf{A}^{\!\mathrm{p}}
    \end{bmatrix}
    \begin{bmatrix}
        \mbf{x}_{k}^{\mathrm{c}} \\
        \mbs{\vartheta}_{k}^{\mathrm{p}}
    \end{bmatrix}
    +
    \begin{bmatrix}
        \mbf{B}^{\mathrm{c}}
        &
        \mbf{0}
        \\
        \mbf{B}^{\mathrm{p}}
        \mbf{D}^{\mathrm{c}}
        &
        \mbf{B}^{\mathrm{p}}
    \end{bmatrix}
    \begin{bmatrix}
        \mbf{u}_{k}^{\mathrm{c}}\\
        \mbf{f}_{k}
    \end{bmatrix},
    \\
    \mbs{\zeta}_{k}^{\mathrm{p}}
    &=
    \begin{bmatrix}
        \mbf{D}^{\mathrm{p}} \mbf{C}^{\mathrm{c}} &
        \mbf{C}^{\mathrm{p}}
    \end{bmatrix}
    \begin{bmatrix}
        \mbf{x}_{k}^{\mathrm{c}} \\
        \mbs{\vartheta}_{k}^{\mathrm{p}}
    \end{bmatrix}
    +
    \begin{bmatrix}
        \mbf{D}^{\mathrm{p}}
        \mbf{D}^{\mathrm{c}}
        &
        \mbf{D}^{\mathrm{p}}
    \end{bmatrix}
    \begin{bmatrix}
        \mbf{u}_{k}^{\mathrm{c}}\\
        \mbf{f}_{k}
    \end{bmatrix},
\end{align}
or equivalently,
\begin{align}
    \mbs{\vartheta}_{k+1}
    &=
    \begin{bmatrix}
        \mbf{A}^{\!\mathrm{c}} &
        \mbf{0} \\
        \mbf{B}^{\mathrm{p}} \mbf{C}^{\mathrm{c}} &
        \mbf{A}^{\!\mathrm{p}}
    \end{bmatrix}
    \mbs{\vartheta}_{k}
    +
    \begin{bmatrix}
        \mbf{B}^{\mathrm{c}}
        &
        \mbf{0}
        \\
        \mbf{B}^{\mathrm{p}}
        \mbf{D}^{\mathrm{c}}
        &
        \mbf{B}^{\mathrm{p}}
    \end{bmatrix}
    \mbs{\upsilon}_{k}^{\mathrm{s}},\label{eq:open_loop_state}
    \\
    \mbs{\zeta}_{k}
    &=
    \begin{bmatrix}
        \mbf{D}^{\mathrm{p}} \mbf{C}^{\mathrm{c}} &
        \mbf{C}^{\mathrm{p}}
    \end{bmatrix}
    \mbs{\vartheta}_{k}
    +
    \begin{bmatrix}
        \mbf{D}^{\mathrm{p}}
        \mbf{D}^{\mathrm{c}}
        &
        \mbf{D}^{\mathrm{p}}
    \end{bmatrix}
    \mbs{\upsilon}_{k}^{\mathrm{s}}.\label{eq:open_loop_output}
\end{align}

\begin{figure}[htb]
    \centering
    \begin{tikzpicture}

        \node[
            smallblock
        ] (plant) at (0, 0) {$\mbs{\mc{P}}$};

        \node[
            summ,
            left = 3ex of plant,
        ] (sum) {};

        \node[
            smallblock,
            left=3ex of sum,
        ] (cont_i) {$\mbs{\mc{C}}$};

        \draw[-stealth] (cont_i) -- (sum);

        \draw[-stealth] (sum) -- (plant);

        \node[
            diff,
            left = 4ex of cont_i
        ] (diff_i) {};
        \draw[-stealth] (diff_i) -- (cont_i)
            node[midway,above] {$\mbf{u}_k^{\mathrm{c}}$};
        \draw[-stealth] (plant.east) -| ++(2ex, -7ex) -| (diff_i)
            node[near end, right]{$\mbs{\zeta}_k$};

        \node[
            left = 4ex of diff_i
        ] (i_bar) {$\mbf{r}_k$};

        \draw[-stealth] (i_bar) -- (diff_i);

        \node[
            above = 3ex of i_bar
        ] (ff) {$\mbf{f}_k$};

        \draw[-stealth] (ff) -| (sum);

        \draw[
            decorate,
            decoration = {brace},
            line width = 1pt,
        ] ([xshift=-2ex]i_bar.south) -- ([xshift=-2ex]ff.north)
        node[
            midway, left
        ] {$\mbs{\upsilon}_k=$};

        \node[
            right = 4ex of plant,
        ] (out) {$\mbs{\zeta}_k$};
        \draw[-stealth] (plant) -- (out);

    \end{tikzpicture}
    \caption{Feedback interconnection of the controller and
    plant.}\label{fig:feedback_interconnection}
\end{figure}
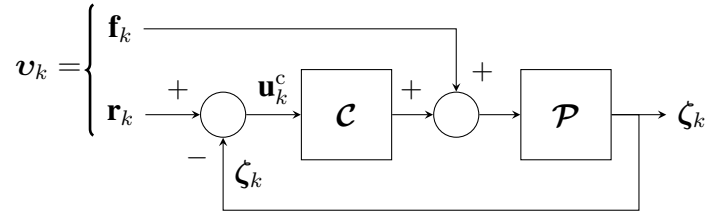

Next, consider the negative feedback interconnection,
\begin{equation}
    \mbf{u}_k^{\mathrm{c}}
    =
    \mbf{r}_k
    -
    \mbs{\zeta}_k,\label{eq:feedback_u}
\end{equation}
where $\mbf{r}_k$ is an exogenous reference signal. The input of
this new feedback-interconnected system is defined to be
\begin{equation}
    \mbs{\upsilon}_k
    =
    \begin{bmatrix}
        \mbf{r}_k \\
        \mbf{f}_k
    \end{bmatrix},
\end{equation}
as depicted in Figure~\ref{fig:feedback_interconnection}.
Substituting~\eqref{eq:feedback_u} into~\eqref{eq:upsilon_s}
yields
\begin{align}
    \mbs{\upsilon}_{k}^{\mathrm{s}}
    &=
    \begin{bmatrix}
        \mbf{r}_k - \mbs{\zeta}_k \\
        \mbf{f}_{k}
    \end{bmatrix}\label{eq:feedback_feedforward}
    \\
    &=
    \mbs{\upsilon}_{k}
    -
    \begin{bmatrix}
        \mbs{\zeta}_k \\
        \mbf{0}
    \end{bmatrix}.
    \label{eq:feedback}
\end{align}

Substituting~\eqref{eq:feedback} into~\eqref{eq:open_loop_output} results in a
new output equation,
\begin{equation}
    \mbs{\zeta}_{k}
    =
    {\left(
        \mbf{1}
        +
        \mbf{D}^{\mathrm{p}}
        \mbf{D}^{\mathrm{c}}
    \right)}^{-1}
    \begin{bmatrix}
        \mbf{D}^{\mathrm{p}} \mbf{C}^{\mathrm{c}} &
        \mbf{C}^{\mathrm{p}}
    \end{bmatrix}
    \mbs{\vartheta}_{k}
    +
    {\left(
        \mbf{1}
        +
        \mbf{D}^{\mathrm{p}}
        \mbf{D}^{\mathrm{c}}
    \right)}^{-1}
    \begin{bmatrix}
        \mbf{D}^{\mathrm{p}}
        \mbf{D}^{\mathrm{c}} &
        \mbf{D}^{\mathrm{p}}
    \end{bmatrix}
    \mbs{\upsilon}_k,\label{eq:feedback_output}
\end{equation}
where the feedback interconnection is \textit{well-posed} if and only if
${\mbf{1} + \mbf{D}^{\mathrm{p}} \mbf{D}^{\mathrm{c}}}$ is
invertible~\cite[\S4.9.1]{skogestad_2006_multivariable}.
Let ${\mbf{Q} = \mbf{1} + \mbf{D}^{\mathrm{p}} \mbf{D}^{\mathrm{c}}}$.
Substituting~\eqref{eq:feedback_output}
into~\eqref{eq:feedback} yields
\begin{equation}
    \mbs{\upsilon}_k^{\mathrm{s}}
    =
    \begin{bmatrix}
        -{\mbf{Q}}^{-1}
        \mbf{D}^{\mathrm{p}} \mbf{C}^{\mathrm{c}} &
        -{\mbf{Q}}^{-1}
        \mbf{C}^{\mathrm{p}} \\
        \mbf{0} &
        \mbf{0}
    \end{bmatrix}
    \mbs{\vartheta}_{k}
    +
    \begin{bmatrix}
        \mbf{1}
        -
        {\mbf{Q}}^{-1}
        \mbf{D}^{\mathrm{p}}
        \mbf{D}^{\mathrm{c}} &
        -
        {\mbf{Q}}^{-1}
        \mbf{D}^{\mathrm{p}} \\
        \mbf{0} &
        \mbf{1}
    \end{bmatrix}
    \mbs{\upsilon}_k.\label{eq:feedback_input}
\end{equation}
Substituting~\eqref{eq:feedback_input} back into~\eqref{eq:open_loop_state}
and rearranging results in a new state equation,
\begin{small}\begin{equation}
    \mbs{\vartheta}_{k+1}
    =
    \begin{bmatrix}
        \mbf{A}^{\!\mathrm{c}}
            - \mbf{B}^{\mathrm{c}}
            {\mbf{Q}}^{-1}
            \mbf{D}^{\mathrm{p}}\mbf{C}^{\mathrm{c}}
            &
        - \mbf{B}^{\mathrm{c}}
            {\mbf{Q}}^{-1}
            \mbf{C}^{\mathrm{p}}
            \\
        \mbf{B}^{\mathrm{p}} \mbf{C}^{\mathrm{c}}
            - \mbf{B}^{\mathrm{p}} \mbf{D}^{\mathrm{c}}
            {\mbf{Q}}^{-1}
            \mbf{D}^{\mathrm{p}}\mbf{C}^{\mathrm{c}}
            &
        \mbf{A}^{\!\mathrm{p}}
            - \mbf{B}^{\mathrm{p}} \mbf{D}^{\mathrm{c}}
            {\mbf{Q}}^{-1}
            \mbf{C}^{\mathrm{p}}
            \\
    \end{bmatrix}
    \mbs{\vartheta}_{k}
    +
    \begin{bmatrix}
        \mbf{B}^{\mathrm{c}} - \mbf{B}^{\mathrm{c}}
            {\mbf{Q}}^{-1}
            \mbf{D}^{\mathrm{p}}
            \mbf{D}^{\mathrm{c}}
        &
        -
        \mbf{B}^{\mathrm{c}}
        {\mbf{Q}}^{-1}
        \mbf{D}^{\mathrm{p}}
        \\
        \mbf{B}^{\mathrm{p}} \mbf{D}^{\mathrm{c}}
            - \mbf{B}^{\mathrm{p}} \mbf{D}^{\mathrm{c}}
            {\mbf{Q}}^{-1}
            \mbf{D}^{\mathrm{p}}
            \mbf{D}^{\mathrm{c}}
        &
        \mbf{B}^{\mathrm{p}}
        -
        \mbf{B}^{\mathrm{p}}
        \mbf{D}^{\mathrm{c}}
        {\mbf{Q}}^{-1}
        \mbf{D}^{\mathrm{p}}
    \end{bmatrix}
    \mbs{\upsilon}_{k},
\end{equation}\end{small}
and the output equation being
\begin{equation}
    \mbs{\zeta}_{k}
    =
    \begin{bmatrix}
        {\mbf{Q}}^{-1}
        \mbf{D}^{\mathrm{p}} \mbf{C}^{\mathrm{c}} &
        {\mbf{Q}}^{-1}
        \mbf{C}^{\mathrm{p}}
    \end{bmatrix}
    \mbs{\vartheta}_{k}
    +
    \begin{bmatrix}
        {\mbf{Q}}^{-1}
        \mbf{D}^{\mathrm{p}}
        \mbf{D}^{\mathrm{c}} &
        {\mbf{Q}}^{-1}
        \mbf{D}^{\mathrm{p}}
    \end{bmatrix}
    \mbs{\upsilon}_k.
\end{equation}

In the Koopman system, only $\mbf{A}^{\!\mathrm{p}}$ and $\mbf{B}^{\mathrm{p}}$
are determined from experimental data. The remaining state-space matrices are
chosen to be
\begin{align}
    \mbf{C}^{\mathrm{p}}
    &=
    \begin{bmatrix}
        \mbf{1}_{m \times m} & \mbf{0}_{m \times (p_{\vartheta} - m)}
    \end{bmatrix},\label{eq:C_def}
    \\
    \mbf{D}^{\mathrm{p}}
    &=
    \mbf{0},\label{eq:D_def}
\end{align}
such that $\mbf{C}^{\mathrm{p}}$ recovers the controlled states of the nonlinear
system being modelled, which are the first $m$ states in
$\mbs{\vartheta}_k^\mathrm{p}$.
Substituting in~\eqref{eq:D_def} implies that ${\mbf{Q} = \mbf{1}}$. Thus, the
simplified state-space representation of the closed-loop system is
\begin{align}
    \mbs{\vartheta}_{k+1}
    &=
    \underbrace{%
        \begin{bmatrix}
            \mbf{A}^{\!\mathrm{c}} &
            - \mbf{B}^{\mathrm{c}} \mbf{C}^{\mathrm{p}} \\
            \mbf{B}^{\mathrm{p}} \mbf{C}^{\mathrm{c}} &
            \mbf{A}^{\!\mathrm{p}} - \mbf{B}^{\mathrm{p}} \mbf{D}^{\mathrm{c}} \mbf{C}^{\mathrm{p}} \\
        \end{bmatrix}
    }_{\mbf{A}^{\!\mathrm{f}}}
    \mbs{\vartheta}_{k}
    +
    \underbrace{%
        \begin{bmatrix}
            \mbf{B}^{\mathrm{c}}
            &
            \mbf{0}
            \\
            \mbf{B}^{\mathrm{p}} \mbf{D}^{\mathrm{c}}
            &
            \mbf{B}^{\mathrm{p}}
        \end{bmatrix}
    }_{\mbf{B}^\mathrm{f}}
    \mbs{\upsilon}_{k},
    \\
    \mbs{\zeta}_{k}
    &=
    \underbrace{%
        \begin{bmatrix}
            \mbf{0} &
            \mbf{C}^{\mathrm{p}}
        \end{bmatrix}
    }_{\mbf{C}^\mathrm{f}}
    \mbs{\vartheta}_{k}
    +
    \underbrace{%
        \mbf{0}
    }_{\mbf{D}^\mathrm{f}}
    \,
    \mbs{\upsilon}_{k},
\end{align}
which is always a well-posed feedback interconnection, since
${\mbf{Q} = \mbf{1} + \mbf{D}^{\mathrm{p}} \mbf{D}^{\mathrm{c}} = \mbf{1}}$
is invertible.

\subsection{Identification of the closed-loop and plant systems}

Consider the closed-loop lifted dataset
${\mathcal{D} = {\left\{\mbs{\vartheta}_k, \mbs{\upsilon}_k\right\}}_{k=0}^q}$,
where the matrix approximation of the Koopman operator is
\begin{align}
    \mbf{U}^{\mathrm{f}}
    &=
    \begin{bmatrix}
        \mbf{U}_{11}^{\mathrm{f}} &
        \mbf{U}_{12}^{\mathrm{f}} &
        \mbf{U}_{13}^{\mathrm{f}} &
        \mbf{U}_{14}^{\mathrm{f}} \\
        \mbf{U}_{21}^{\mathrm{f}} &
        \mbf{U}_{22}^{\mathrm{f}} &
        \mbf{U}_{23}^{\mathrm{f}} &
        \mbf{U}_{24}^{\mathrm{f}}
    \end{bmatrix}.\label{eq:Uf_block}
    \\
    &=
    \begin{bmatrix}
        \mbf{A}^{\!\mathrm{c}} &
        - \mbf{B}^{\mathrm{c}} \mbf{C}^{\mathrm{p}} &
        \mbf{B}^{\mathrm{c}} &
        \mbf{0}
        \\
        \mbf{B}^{\mathrm{p}} \mbf{C}^{\mathrm{c}} &
        \mbf{A}^{\!\mathrm{p}} - \mbf{B}^{\mathrm{p}} \mbf{D}^{\mathrm{c}} \mbf{C}^{\mathrm{p}} &
        \mbf{B}^{\mathrm{p}} \mbf{D}^{\mathrm{c}} &
        \mbf{B}^{\mathrm{p}}
    \end{bmatrix}\label{eq:Uf_relation}
\end{align}
Comparing~\eqref{eq:Uf_block} and~\eqref{eq:Uf_relation} reveals that
the matrices, $\mbf{B}^\mathrm{p}$, $\mbf{U}_{21}^{\mathrm{f}}$,
$\mbf{U}_{23}^{\mathrm{f}}$, and $\mbf{U}_{24}^{\mathrm{f}}$ are
related via the expression
\begin{equation}
    \mbf{B}^{\mathrm{p}}
    {\begin{bmatrix}
        \mbf{C}^{\mathrm{c}} &
        \mbf{D}^{\mathrm{c}} &
        \mbf{1}
        \vphantom{\mbf{U}_{24}^{\mathrm{f}}}
    \end{bmatrix}}
    =
    \begin{bmatrix}
        \mbf{U}_{21}^{\mathrm{f}} &
        \mbf{U}_{23}^{\mathrm{f}} &
        \mbf{U}_{24}^{\mathrm{f}}
    \end{bmatrix}.\label{eq:constr_Bp}
\end{equation}
Thus, one way to compute
$\mbf{U}^{\mathrm{p}}
    =
    \begin{bmatrix}
    \mbf{A}^{\!\mathrm{p}} &
    \mbf{B}^{\mathrm{p}}
\end{bmatrix}$
is to first compute $\mbf{U}^{\mathrm{f}}$ using EDMD and then recover the
plant's state space matrices using
\begin{align}
    \mbf{B}^{\mathrm{p}}
    &=
    \begin{bmatrix}
        \mbf{U}_{21}^{\mathrm{f}} &
        \mbf{U}_{23}^{\mathrm{f}} &
        \mbf{U}_{24}^{\mathrm{f}}
    \end{bmatrix}
    {\begin{bmatrix}
        \mbf{C}^{\mathrm{c}} &
        \mbf{D}^{\mathrm{c}} &
        \mbf{1}
        \vphantom{\mbf{U}_{24}^{\mathrm{f}}}
    \end{bmatrix}}^\dagger,\label{eq:lstsq_Bp}
    \\
    \mbf{A}^{\!\mathrm{p}}
    &=
    \mbf{U}_{22}^{\mathrm{f}}
    +
    \mbf{B}^{\mathrm{p}}
    \mbf{D}^{\mathrm{c}}
    \mbf{C}^{\mathrm{p}}.\label{eq:constr_Ap}
\end{align}
While this approach is simple, it is not guaranteed to preserve the required
relationships between $\mbf{U}^{\mathrm{f}}$, $\mbf{A}^{\!\mathrm{p}}$, and
$\mbf{B}^{\mathrm{c}}$ found in~\eqref{eq:Uf_block} and~\eqref{eq:Uf_relation}.
As a result, identifying the closed-loop system, extracting the plant model
with~\eqref{eq:lstsq_Bp} and~\eqref{eq:constr_Ap}, and then closing the loop
again with the same controller using~\eqref{eq:Uf_relation} can result in an
entirely different closed-loop system. This pitfall is explored further in the
\hyperref[sec:lstsq]{Appendix}.

To preserve the structure of the closed-loop system,~\eqref{eq:constr_Bp}
and~\eqref{eq:constr_Ap} can be treated as constraints when identifying the
Koopman matrix of the closed-loop system. Including Tikhonov regularization, the
resulting EDMD optimization problem is
\begin{align}
    \min\;&
    J(\mbf{U}^{\mathrm{f}}, \mbf{U}^{\mathrm{p}}; \alpha)
    =
    \frac{1}{q}
    \big\|\mbs{\Theta}_+ - \mbf{U}^{\mathrm{f}}\mbs{\Psi}\big\|_\frob^2
    +
    \frac{\alpha}{q}
    \big\|\mbf{U}^{\mathrm{f}}\big\|_\frob^2\label{eq:opt_abstract_cost}
    \\
    \mathrm{s.t.}\;&
    \begin{bmatrix}
        \mbf{U}_{21}^{\mathrm{f}} &
        \mbf{U}_{23}^{\mathrm{f}} &
        \mbf{U}_{24}^{\mathrm{f}}
    \end{bmatrix}
    =
    \mbf{B}^{\mathrm{p}}
    \begin{bmatrix}
        \mbf{C}^{\mathrm{c}} &
        \mbf{D}^{\mathrm{c}} &
        \mbf{1}
    \end{bmatrix},\label{eq:opt_abstract_Bp}
    \\
    &\mbf{U}_{22}^{\mathrm{f}}
    =
    \mbf{A}^{\!\mathrm{p}}
    -
    \mbf{B}^{\mathrm{p}}
    \mbf{D}^{\mathrm{c}}
    \mbf{C}^{\mathrm{p}}.\label{eq:opt_abstract_Ap}
\end{align}

To efficiently solve the optimization problem posed
in~\eqref{eq:opt_abstract_cost}--\eqref{eq:opt_abstract_Ap}, its cost function
is reformulated as a semidefinite program in a manner similar
to~\cite{dahdah_linear_2021, dahdah_system_2022}.
The cost function in~\eqref{eq:opt_abstract_cost} can be rewritten as
\begin{equation}
    J(\mbf{U}^{\mathrm{f}}; \alpha)
    =
    \frac{1}{q}
    \trace{\left(
        \mbs{\Theta}_+
        \mbs{\Theta}_+^\trans
        -
        \He{
            \mbf{U}^{\mathrm{f}}
            \mbs{\Psi}
            \mbs{\Theta}_+^\trans
        }
        +
        \mbf{U}^{\mathrm{f}}
        \mbs{\Psi}
        \mbs{\Psi}^\trans
        \mbf{U}^{\mathrm{f}^\trans}
    \right)}
    \\
    +
    \frac{\alpha}{q}
    \trace{\left(
        \mbf{U}^{\mathrm{f}}
        \mbf{U}^{\mathrm{f}^\trans}
    \right)},
\end{equation}
where $\He{\cdot} = (\cdot) + {(\cdot)}^\trans$. Substituting
in~\eqref{eq:edmdi-sol-scaled} yields
\begin{equation}
    J(\mbf{U}^{\mathrm{f}}; \alpha)
    =
    \trace{\left(
        \mbf{F}
        -
        \He{
            \mbf{U}^{\mathrm{f}}
            \mbf{G}^\trans
        }
        +
        \mbf{U}^{\mathrm{f}}
        \mbf{H}_\alpha
        \mbf{U}^{\mathrm{f}^\trans}
    \right)},
\end{equation}
where $\mbf{F} = \frac{1}{q} \mbs{\Theta}_+ \mbs{\Theta}_+^\trans$
and
$\mbf{H}_\alpha = \mbf{H} + \frac{\alpha}{q}\mbf{1}$.
Introducing a slack variable~\cite[\S2.15.1]{caverly_2019_lmi} results in the
equivalent optimization problem,
\begin{align}
    \min\;&
    J(\mbf{U}^{\mathrm{f}}, \mbf{W}; \alpha)
    =
    \trace{\left(\mbf{W}\right)}
    \\
    \mathrm{s.t.}\;&
    \mbf{W} > 0,
    \\
    &\mbf{F}
    -
    \He{
        \mbf{U}^{\mathrm{f}}
        \mbf{G}^\trans
    }
    +
    \mbf{U}^{\mathrm{f}}
    \mbf{H}_\alpha
    \mbf{U}^{\mathrm{f}^\trans}
    <
    \mbf{W}.\label{eq:constr-pre-schur}
\end{align}
Next, the Schur complement is used to break up the quadratic term
in~\eqref{eq:constr-pre-schur}~\cite[\S2.3.1]{caverly_2019_lmi}.
Equation~\eqref{eq:edmdi-sol-scaled} shows that ${\mbf{H} = \mbf{H}^\trans > 0}$
if the columns of $\mbs{\Psi}$ are linearly independent. Assuming that
$\mbf{H}_\alpha$ is positive definite, consider its Cholesky factorization,
$\mbf{H}_\alpha = \mbf{R}_\alpha \mbf{R}_\alpha^\trans$. Substituting in the
Cholesky factorization and applying the Schur complement yields the semidefinite
program,
\begin{align}
    \min\;&
    J(\mbf{U}^{\mathrm{f}}, \mbf{W}; \alpha)
    =
    \trace\left(\mbf{W}\right)
    \\
    \mathrm{s.t.}\;&
    \mbf{W} > 0,
    \\
    &\begin{bmatrix}
        -\mbf{W} + \mbf{F} - \He{\mbf{U}^{\mathrm{f}} \mbf{G}^\trans} &
        \mbf{U}^{\mathrm{f}} \mbf{R}_\alpha \\
        \mbf{R}_\alpha^\trans \mbf{U}^{\mathrm{f}^\trans} &
        -\mbf{1}
    \end{bmatrix} < 0.
\end{align}
Including the structural constraints on the closed-loop system
from~\eqref{eq:constr_Bp} and~\eqref{eq:constr_Ap} results in
\begin{align}
    \min\;&
    J(\mbf{U}^{\mathrm{f}}, \mbf{U}^{\mathrm{p}}, \mbf{W}; \alpha)
    =
    \trace\left(\mbf{W}\right)
    \\
    \mathrm{s.t.}\;&
    \mbf{W} > 0,
    \\
    &\begin{bmatrix}
        -\mbf{W} + \mbf{F} - \He{\mbf{U}^{\mathrm{f}} \mbf{G}^\trans} &
        \mbf{U}^{\mathrm{f}} \mbf{R}_\alpha \\
        \mbf{R}_\alpha^\trans \mbf{U}^{\mathrm{f}^\trans} &
        -\mbf{1}
    \end{bmatrix} < 0,
    \\
    &\begin{bmatrix}
        \mbf{U}_{21}^{\mathrm{f}} &
        \mbf{U}_{23}^{\mathrm{f}} &
        \mbf{U}_{24}^{\mathrm{f}}
    \end{bmatrix}
    =
    \mbf{B}^{\mathrm{p}}
    \begin{bmatrix}
        \mbf{C}^{\mathrm{c}} &
        \mbf{D}^{\mathrm{c}} &
        \mbf{1}
    \end{bmatrix},
    \\
    &\mbf{U}_{22}^{\mathrm{f}}
    =
    \mbf{A}^{\!\mathrm{p}}
    -
    \mbf{B}^{\mathrm{p}}
    \mbf{D}^{\mathrm{c}}
    \mbf{C}^{\mathrm{p}},
\end{align}
which provides a means to simultaneously identify a Koopman model of the
closed-loop system and the plant system.
While this method applies Tikhonov regularization to the closed-loop system's
Koopman matrix, any regularizer or constraint, such as those
in~\cite{dahdah_linear_2021, dahdah_system_2022},
can be applied to either the closed-loop system or the plant system, since both
Koopman matrices are optimization variables.

\subsection{Discussion of bias in EDMD}
Even when identifying a plant operating in open-loop, EDMD is known to identify
biased Koopman operators when sensor noise is
significant~\cite[\S2.2]{dawson_2016_characterizing}. However, since the plant's
input is exogenous in an open-loop setting, any noise in the input signal is
uncorrelated with the measurement noise. In fact, the input signal is often
known exactly because it is chosen by the user.

In contrast, consider a scenario where measurements are gathered from a system
operating in closed-loop with a controller. Feedback action propagates noise
from the plant's measured output throughout the system.
If the presence of the controller is ignored and its control signal to the plant
is treated as exogenous, correlations between the plant input and measurement
noise result in increased bias~\cite{forssell_1999_closedloop,
    hof_1998_closedloop},~\cite[\S11.1]{katayama_2005_subspace}. Furthermore, the
input signal is no longer known exactly, as it is corrupted by noise that is
correlated with the state's measurement noise.

Identifying the Koopman system using a closed-loop procedure sidesteps this
issue by using the exogenous reference and feedforward signals as inputs. If
these signals are noisy, the noise is uncorrelated with the state's measurement
noise. However, these inputs are typically known exactly. Using a closed-loop
approach therefore reduces bias compared to treating the feedback controller's
output as an exogenous plant input.
The remaining bias, which is due solely to measurement noise, is inherent
to EDMD.\@ In practice, this bias is sometimes too small to justify the use of
more complex identification techniques~\cite[\S11.1]{katayama_2005_subspace}.
Variants of DMD like forward-backward
DMD~\cite[\S2.4]{dawson_2016_characterizing} and total least-squares
DMD~\cite[\S2.5]{dawson_2016_characterizing} can also be used to compensate for
this bias.

\section{Simulated example}\label{sec:duffing_example}

{Several advantages of the proposed closed-loop Koopman operator
approximation method, referred to in this section as closed-loop EDMD
(CL~EDMD), are demonstrated on a simulated closed-loop Duffing oscillator
system, pictured in Figure~\ref{fig:duffing}.

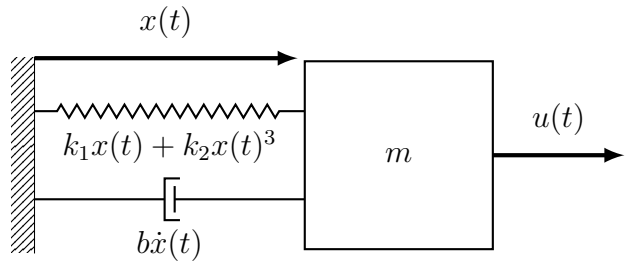
\begin{figure}[htb]
    \centering
    \begin{tikzpicture}
        \tikzstyle{spring}=[
            thick,
            decorate,
            decoration={zigzag,pre length=0.3cm,post length=0.3cm,segment length=6},
        ]
        \tikzstyle{damper}=[
            thick,
            decoration={
                markings,  
                mark connection node=dmp,
                mark=at position 0.5 with {
                  \node (dmp) [thick,inner sep=0pt,transform
                  shape,rotate=-90,minimum width=15pt,minimum height=3pt,draw=none]
                  {};
                  \draw [thick] ($(dmp.north east)+(2pt,0)$) -- (dmp.south east) --
                  (dmp.south west) -- ($(dmp.north west)+(2pt,0)$);
                  \draw [thick] ($(dmp.north)+(0,-5pt)$) --
                  ($(dmp.north)+(0,5pt)$);
                }
            },
            decorate,
        ]
        \tikzstyle{ground}=[
            fill,
            pattern=north east lines,
            draw=none,
            minimum width=0.75cm,
            minimum height=0.3cm
        ]

        \node (M) [
            draw,
            outer sep=0pt,
            thick,
            minimum width=2.5cm,
            minimum height=2.5cm,
        ] {$m$};
        \node (wall) [ground, rotate=-90, minimum width=2.6cm,yshift=-5cm] {};

        \draw (wall.north east) -- (wall.north west);
        \draw [spring] (wall.165) -- ($(M.north west)!(wall.165)!(M.south west)$)
            node[midway, below=0.10cm] {$k_1 x(t) + k_2 x(t)^3$};
        \draw [damper] (wall.15) -- ($(M.north west)!(wall.15)!(M.south west)$)
            node[midway, below=0.25cm] {$b \dot{x}(t)$};
        \draw [-latex, ultra thick] (wall.170) ++(0,0.4cm) -- ++(3.5cm,0)
            node[midway, above=0.10cm] {$x(t)$};
        \draw [-latex, ultra thick] (M) -- ++(3cm,0)
            node[midway, above=0.10cm] {$u(t)$};
    \end{tikzpicture}
    \caption{Duffing oscillator system with position~$x(t)$, force~$u(t)$,
    mass~$m$, viscous damping~$b$, linear stiffness~$k_1$, and nonlinear
    stiffness~$k_2$.}\label{fig:duffing}
\end{figure}

\subsection{Simulation setup}
The Duffing oscillator~\cite{duffing} can be viewed as a forced
mass-spring-damper system with nonlinear spring stiffness. That is,
\begin{equation}
    m \ddot{x}(t) + b \dot{x}(t) + k_1 x(t) + k_2 {x(t)}^3 = u(t),
\end{equation}
where
$x(t)$ is the mass position (\si{\metre}), $u(t)$ is the external force
(\si{\newton}).
A simulated dataset is generated using a Duffing oscillator with mass
$m=\SI{0.01}{\kilo\gram}$, viscous damping
$b=\SI{0.1}{\si{\newton\second\per\metre}}$, linear stiffness
$k_1=\SI{0.02}{\si{\newton\per\metre}}$, and nonlinear stiffness
$k_2=\SI{0.4}{\si{\newton\per\metre^3}}$.
Its position error $e(t) = r(t) - x(t)$ is controlled by a
proportional-integral controller,
\begin{equation}
    K(s) = K^\mathrm{p} + \frac{1}{s} K^\mathrm{i},
\end{equation}
with coefficients
$K^\mathrm{p}=\SI{1}{\newton\per\metre}$ and
$K^\mathrm{i}=\SI{1}{\newton\per\metre\per\second}$. Thus, following the
notation of Section~\ref{sec:closed_loop}, $x^\mathrm{p}(t) = x(t)$,
$\upsilon^\mathrm{p}(t) = u(t)$, and $\upsilon(t) = r(t)$.

The simulated dataset consists of 10 training episodes and one test
episode. Each episode is \SI{10}{\second} long and begins with zero initial
conditions. The episodes are sampled at a frequency of \SI{100}{\hertz}. The
position reference is a pseudorandom binary
sequence~\cite[\S13.3]{ljung_1999_system} with amplitude \SI{1}{\metre}. An
example reference signal is pictured in Figure~\ref{fig:duffing_error_cl}.
The training set position measurements used for both control and identification
are subject to additive coloured noise. Noise samples are first drawn from a
Gaussian distribution with zero mean and covariance
$\sigma^2=\SI{0.02}{\metre^2}$. This white noise signal is then filtered using
a 12\textsuperscript{th} order Butterworth filter with a cutoff frequency of
\SI{5}{\hertz}. The test set does not have any added measurement noise.

\subsection{Comparison of identified models}

The Koopman lifting functions used to model the Duffing oscillator
consist of 6 time delays of the state followed by 50 thin plate radial
basis functions (RBFs) of the form
\begin{equation}
    \varphi_i(\mbf{x})
    =
    {\left(\alpha \| \mbf{x} - \mbf{c}_i \| + \delta\right)}^2
    \ln{\left(\alpha \| \mbf{x} - \mbf{c}_i \| + \delta\right)},
    \quad
    i = 1, \ldots, 50,
\end{equation}
where $\alpha=0.2$ is the shape parameter and $\delta=0.001$ is a small offset
to avoid evaluating the logarithm at zero.
The centres $\mbf{c}_i$ are chosen using Latin hypercube sampling.
The same set of lifting functions is used to identify two Koopman models of the
Duffing oscillator: one using EDMD in open-loop, and another using the proposed
CL~EDMD method. In both cases, the inputs are not lifted.
This results in a 57-dimensional lifted state for the open-loop Koopman model,
and a 58-dimensional state for the closed-loop Koopman model, as the
controller's state is also included.

To better understand the advantages and disadvantages of the Koopman
approaches, two autoregressive models with exogenous inputs (ARX
models)~\cite[\S4.2]{ljung_1999_system} are identified using
SIPPY~\cite{sippy}, an open-source system identification library for Python.
For comparison with EDMD, an ARX model of the open-loop plant is identified
using 6 output lags and one input lag, resulting in a 6\textsuperscript{th}
order system. This is an example of the direct approach to closed-loop system
identification.
For comparison with CL~EDMD, an ARX model of the closed-loop system is
identified, also using 6 output lags and one input lag. The transfer function
of the plant is recovered using the indirect approach discussed
in~\cite[\S5.1]{forssell_1999_closedloop} and~\cite[\S13.5]{ljung_1999_system}.
The indirect approach, referred to as CL~ARX in this section, increases the
order of the plant transfer function to 14.
Note that SIPPY only supports identification using a single training episode.
Thus, to make full use of the training data, a transfer function is identified
for each episode and the coefficients are averaged to obtain a final model.

\begin{figure}[htbp]
    \centering
    \begin{subfigure}[t]{3.5in}
        \centering
        \includegraphics[width=\linewidth, trim=0 -0.44cm 0 0]{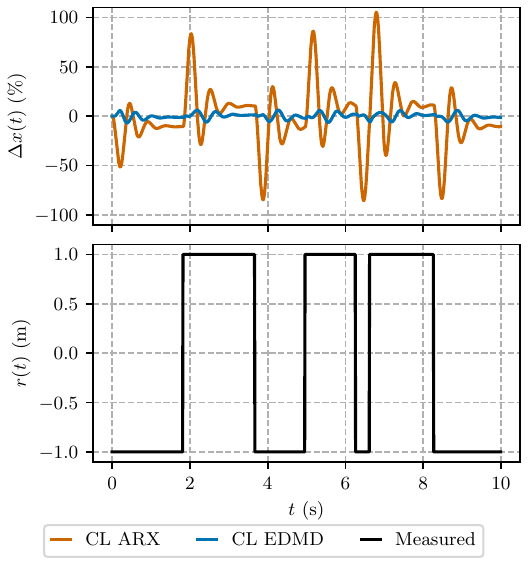}
        \caption{Prediction errors of the closed-loop systems identified using
        CL~ARX and CL~EDMD methods. The Koopman approach better captures the
        dynamics of the closed-loop system.}\label{fig:duffing_error_cl}
    \end{subfigure}%
    \hfill
    \begin{subfigure}[t]{3.5in}
        \centering
        \includegraphics[width=\linewidth]{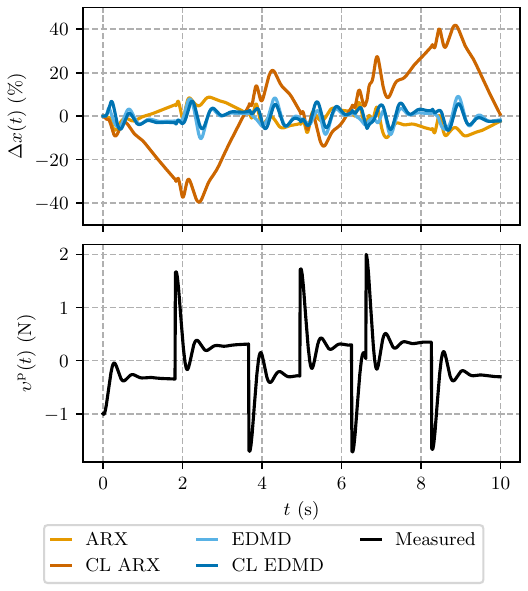}
        \caption{Prediction errors of the plant systems identified using
        closed-loop and open-loop ARX and EDMD methods. The CL~ARX model fails
        to capture the plant dynamics, possibly due to its high order. The
        Koopman models perform best, with nearly identical prediction
        errors.}\label{fig:duffing_error_ol}
    \end{subfigure}
    \caption{Prediction errors the closed-loop and plant systems identified
    using ARX and EDMD methods.}\label{fig:duffing_error}
\end{figure}

\begin{table}[ht]
    \centering
    \caption{Normalized mean and RMS open-loop plant errors in test
    episode.}\label{tab:duffing_score}
    \begin{tabular}{lcc}
        \toprule
        Method &
        Normalized mean error  &
        Normalized RMS error \\
        \midrule
        EDMD & \SI{-0.6}{\percent} & \SI{3.3}{\percent} \\
        \midrule
        CL~EDMD & \SI{-0.1}{\percent} & \SI{3.2}{\percent} \\
        \midrule
        ARX & \SI{-0.9}{\percent} & \SI{4.7}{\percent} \\
        \midrule
        CL~ARX & \SI{2.5}{\percent} & \SI{20.4}{\percent} \\
        \bottomrule
    \end{tabular}
\end{table}

Figure~\ref{fig:duffing_error} shows the prediction errors of all four
models on the test episode. The mean and root-mean-squared (RMS) prediction
errors for the plant system, normalized by the peak amplitude of the state, are
summarized in Table~\ref{tab:duffing_score}.
While the CL~EDMD method is able to identify accurate closed-loop and plant
models, the CL~ARX method is not able to accurately predict the Duffing
oscillator's position throughout the test episode. Its failure to accurately
predict the plant's open-loop dynamics is likely due to its inaccurate
identification of the closed-loop system. Furthermore, the classical indirect
approach to system identification leads to high-order plant transfer
functions~\cite[\S5.1]{forssell_1999_closedloop}, which may impact the accuracy
of the plant model. In contrast, CL~EDMD always identifies a plant system with
fewer states than the closed-loop system, and does not rely on multiplying or
inverting any transfer matrices. While the Koopman models have over 50 states,
only 6 of them are time delays, compared to the 14 of the CL~ARX plant model.
The open-loop ARX model performs better than its closed-loop version, but not
as well as either EDMD or CL~EDMD.\@ Both EDMD methods perform nearly
identically, with CL~EDMD attaining slightly lower mean and RMS errors on the
test episode.

While the simulated Duffing oscillator example clearly demonstrates the
advantages of Koopman approaches to system identification, it does not
highlight the advantages of CL~EDMD over standard EDMD.\@ The unique benefits
of closed-loop EDMD will be explained in Section~\ref{sec:results}, where a
more complex experimental example is considered, and issues such as
hyperparameter optimization and regularization are discussed.

\section{Experimental example}\label{sec:results}
The benefits of the proposed closed-loop EDMD method (CL~EDMD) are demonstrated
on a dataset collected from the Quanser \textit{QUBE-Servo}~\cite{qube}, a
rotary inverted pendulum system. Pictured in Figure~\ref{fig:qube}, this system
has an unstable equilibrium point when the pendulum is upright.
\begin{figure}[htbp]
    \centering
    \includegraphics[width=3.5in, angle=-90]{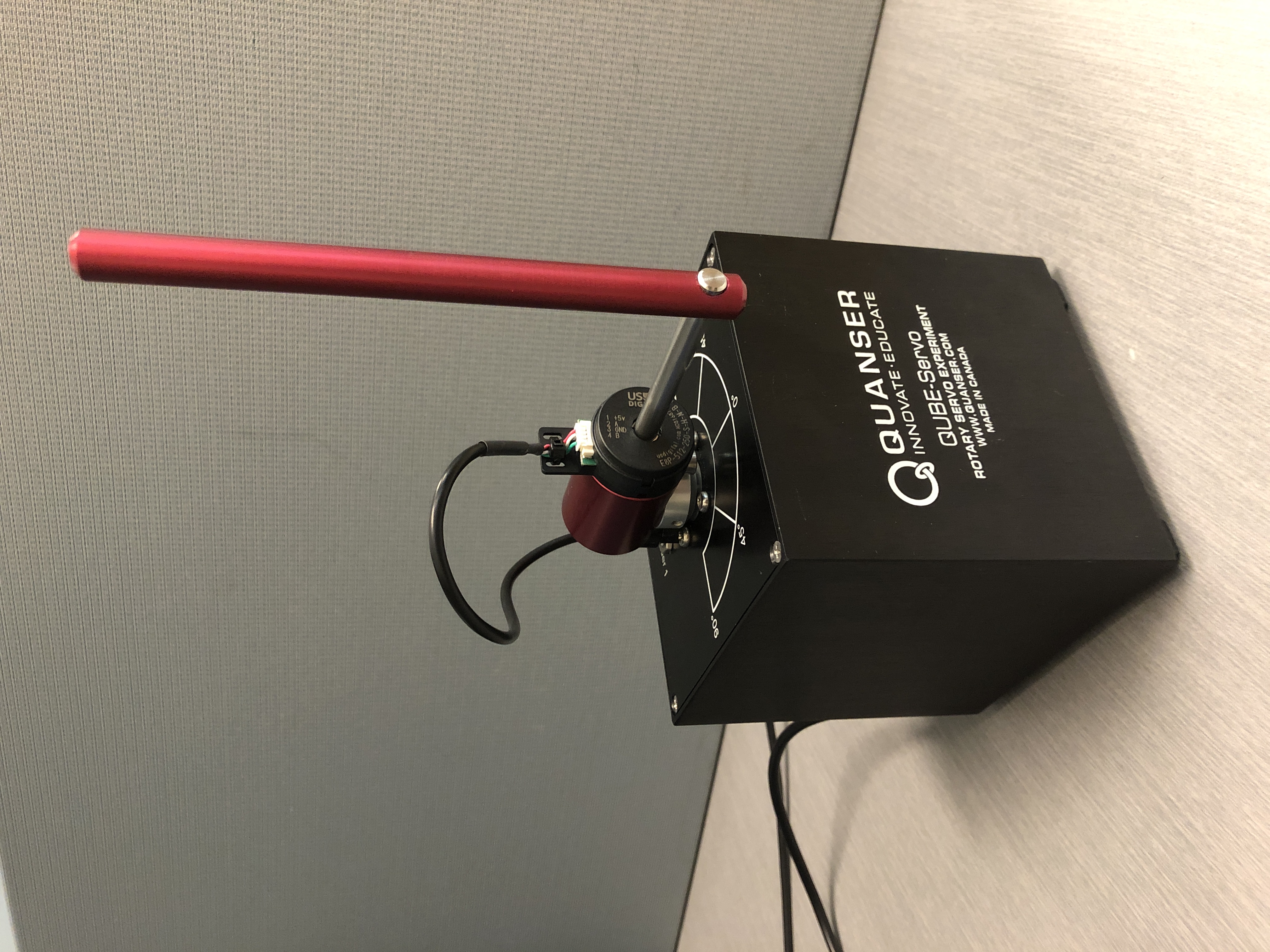}
    \caption{The Quanser \textit{QUBE-Servo} system used to demonstrate the
    proposed closed-loop Koopman operator approximation method.}\label{fig:qube}
\end{figure}
\subsection{Experimental setup}\label{sec:experimental_setup}
The \textit{QUBE-Servo} has one DC motor with a maximum input voltage of
\SI{10}{\volt}, as well as two incremental encoders with \si{2048} counts per
revolution. The system is controlled by two parallel proportional-derivative
controllers that track references for the motor angle $x_1^\mathrm{p}(t)$ and
the pendulum angle $x_2^\mathrm{p}(t)$. The reference signals are denoted
$r_1(t)$ and $r_2(t)$, while the feedforward is denoted $f(t)$. All angles are
presented in radians, while the plant inputs are presented in Volts.

The controller's transfer matrix from tracking error
${\mbf{e}(t) = \mbf{r}(t) - \mbf{x}^\mathrm{p}(t)}$
to output $\upsilon^\mathrm{p}(t)$ is
\begin{equation}
    \mbf{K}(s)
    =
    \begin{bmatrix}
        K_1^\mathrm{p}
        +
        K_1^\mathrm{d}
        \frac{a s}{s + a}
        &
        K_2^\mathrm{p}
        +
        K_2^\mathrm{d}
        \frac{a s}{s + a}
    \end{bmatrix},
\end{equation}
where $K_1^\mathrm{p}=\SI{6}{\volt\per\radian}$ and
$K_2^\mathrm{p}=\SI{30}{\volt\per\radian}$ are the proportional gains,
$K_1^\mathrm{d}=\SI{1.8}{\volt\per\radian}$ and
$K_2^\mathrm{d}=\SI{2.5}{\volt\per\radian}$ are the derivative gains, and
$a=\SI{50}{\hertz}$ is the derivative filter cutoff frequency.
The controller is discretized using the zero-order hold with a sampling
frequency of \SI{500}{\hertz}.

Samples of the exogenous test inputs are pictured in Figure~\ref{fig:inputs_cl}.
The motor angle reference signal is an integrated pseudorandom binary sequence,
while the pendulum angle reference and feedforward signals are pseudorandom
binary sequences~\cite[\S13.3]{ljung_1999_system}. The pseudorandom binary
sequences are smoothed using a low-pass filter with a cutoff frequency of
\SI{200}{\hertz}. The signal amplitudes excite the rotary inverted pendulum as
much as possible without causing the pendulum to fall from its upright position.

\begin{figure}[htb]
    \centering
    \includegraphics[width=3.5in]{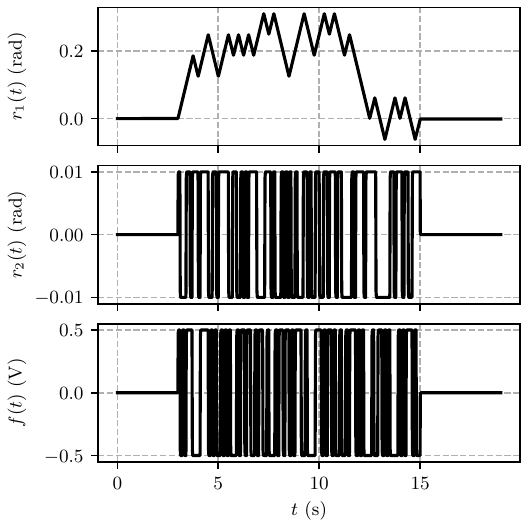}
    \caption{Sample of the exogenous test inputs used to identify a Koopman
    model of the Quanser \textit{QUBE-Servo}.}\label{fig:inputs_cl}
\end{figure}
The dataset presented here consists of \num{30} training episodes and \num{20}
test episodes. Each episode is \SI{20}{\second}, or \num{10000} samples, long.
The motor angle, pendulum angle, motor voltage, reference motor angle, reference
pendulum angle, and feedforward voltage are recorded at every timestep. The
controller's state is computed at each timestep using the recorded tracking
error. The first \SI{1}{\second} of each dataset is discarded to remove the
transients associated with manually raising the pendulum to the upright
position.

The Koopman lifting functions chosen for the rotary inverted pendulum are
second-order monomials followed by \num{10} time delays. As required by the
proposed method, the inputs and controller states are not lifted.
Further performance improvement can be achieved by including more time delays,
with diminishing returns after~\num{20}.

As a point of comparison with classical system identification, a
multiple-input multiple-output (MIMO) ARX model with 10 output lags and 10
input lags is fit to the closed-loop system using SIPPY~\cite{sippy}.
However, a difficulty with the indirect approach to closed-loop system
identification in the MIMO case is that it requires computing the pseudoinverse
of a nonsquare transfer matrix to recover the plant's transfer matrix.
Attempting this symbolically leads to a transfer matrix of an intractable
order. Since the closed-loop system is strictly proper, computing the
pseudoinverse in state-space is also nontrivial. To circumvent this, the $2
\times 1$ block of the closed-loop transfer matrix corresponding to the
feedforward input is used to recover the plant's transfer matrix. This approach
only requires inverting a single-input single-output (SISO) transfer function.

The software required to fully reproduce the results of this paper, including
the \textit{QUBE-Servo} dataset, is available at
\url{https://github.com/decargroup/closed_loop_koopman}. This code extends
\texttt{pykoop}~\cite{pykoop}, the authors' open source Koopman operator
approximation library for Python. The software used to gather the
\textit{QUBE-Servo} dataset, implemented in C, is available at
\url{https://github.com/decargroup/quanser_qube}.

\subsection{Comparison of regularization methods and scoring metrics}\label{sec:regularization}

\begin{figure}[htbp]
    \centering
    \begin{subfigure}[t]{3.5in}
        \centering
        \includegraphics[width=\linewidth]{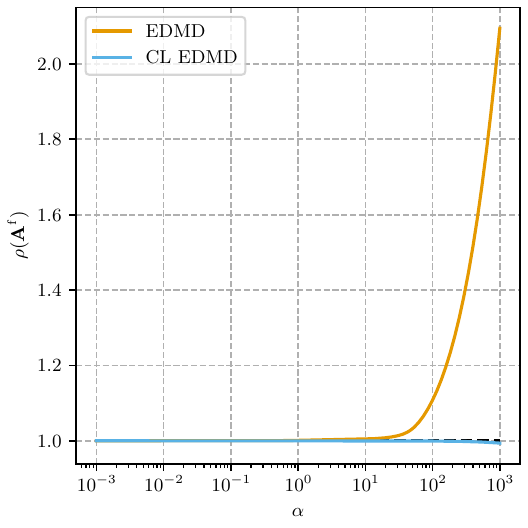}
        \caption{Relationship between the regularization coefficient $\alpha$
        and the spectral radii of the CL systems identified with EDMD
        and CL~EDMD.\@ As $\alpha$ increases, the system identified with EDMD
        becomes unstable, while the one identified with the proposed method
        remains asymptotically stable.}\label{fig:spectral_radius_cl}
    \end{subfigure}%
    \hfill
    \begin{subfigure}[t]{3.5in}
        \centering
        \includegraphics[width=\linewidth]{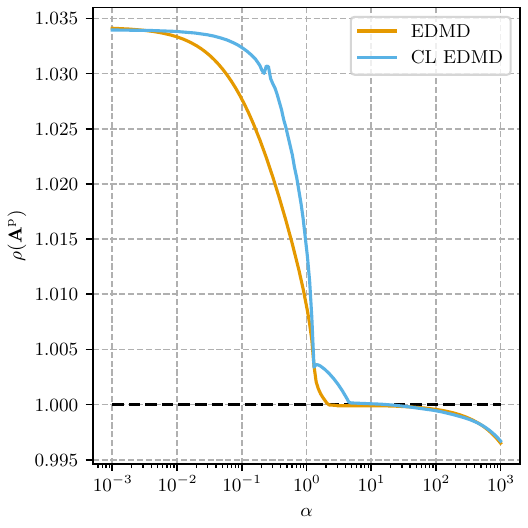}
        \caption{Relationship between the regularization coefficient $\alpha$
        and the spectral radii of the plant systems identified with EDMD and
        CL~EDMD.\@ As $\alpha$ increases, both systems are incorrectly
        stabilized by the regularization.}\label{fig:spectral_radius_ol}
    \end{subfigure}\vspace{2ex}
    \begin{subfigure}[t]{3.5in}
        \centering
        \includegraphics[width=\linewidth]{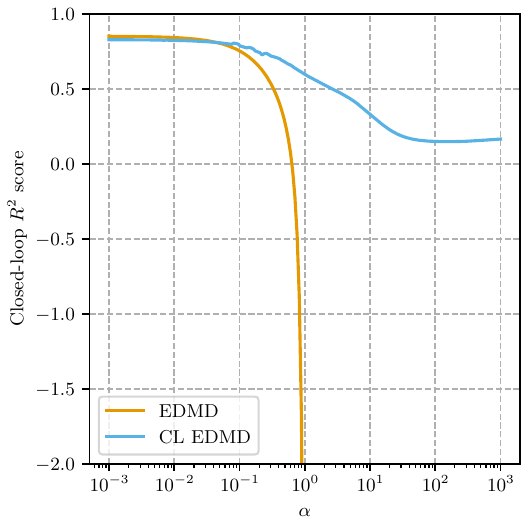}
        \caption{Relationship between the regularization coefficient $\alpha$
        and the test set $R^2$ scores of the CL systems identified with EDMD and
        CL~EDMD.\@ As $\alpha$ increases, both scores decrease. However, the
        system identified with EDMD becomes unstable around $\alpha=1$, and its
        score diverges.}\label{fig:cross_validation_cl}
    \end{subfigure}%
    \hfill
    \begin{subfigure}[t]{3.5in}
        \centering
        \includegraphics[width=\linewidth]{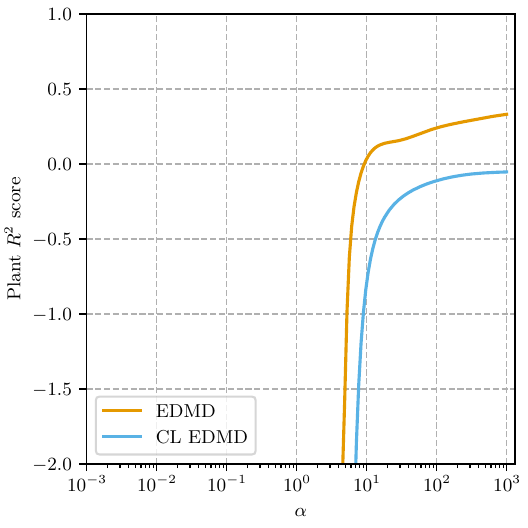}
        \caption{Relationship between the regularization coefficient $\alpha$
        and the test set $R^2$ scores of the plant systems identified with EDMD
        and CL~EDMD.\@ The scores are only finite for large regularization
        coefficients, indicating that is not a good metric for tuning the
        regularizer.}\label{fig:cross_validation_ol}
    \end{subfigure}
    \caption{Effect of regularization on the spectral radii and prediction
    scores of the CL and plant systems.}\label{fig:cross_validation}
\end{figure}

Since the \textit{QUBE-Servo} dataset does not contain significant sensor noise,
Extended DMD does not identify a noticeably biased Koopman model. In fact, with
no regularization, both EDMD and the proposed closed-loop method identify the
same Koopman system.
However, in many situations, regularization is a necessary component of the
Koopman identification process. Choosing an appropriate regularization
coefficient typically requires a hyperparameter optimization procedure.
As such, the advantages of the proposed method are examined through the lens of
hyperparameter optimization.

The proposed closed-loop Koopman operator approximation method is first compared
to EDMD by varying the regularization coefficient $\alpha$ of each
identification method. Specifically, \si{180} values of $\alpha$ are evaluated,
spaced logarithmically between $10^{-3}$ and $10^3$.
Recall that EDMD with Tikhonov regularization penalizes the squared Frobenius
norm of the plant system, $\|\mbf{U}^\mathrm{p}\|_\frob^2$, while the proposed
closed-loop method penalizes the squared Frobenius norm of the closed-loop
system, $\|\mbf{U}^\mathrm{f}\|_\frob^2$.
For the sake of comparison with the closed-loop approach, the open-loop plant
models identified by EDMD are wrapped with the known controller.
Since the proposed approach simultaneously identifies the closed-loop and plant
systems, the plant system identified using the proposed method can be compared
directly to the plant model identified using EDMD.\@ %

As a first point of comparison between EDMD and the proposed closed-loop method,
consider the spectral radii of the closed-loop and plant systems as a
function of their regularization coefficients.
Given prior knowledge of the inverted pendulum system, both in open-loop and
closed-loop contexts, a desirable Koopman model should have asymptotically
stable closed-loop dynamics when controlled using an appropriate controller, but
unstable open-loop dynamics.
Figure~\ref{fig:spectral_radius_cl} shows that regularizing using the Koopman
matrix of the closed-loop system always results in an asymptotically stable
closed-loop system. However, regularizing using the plant's Koopman matrix
results in an unstable closed-loop system for high enough regularization
coefficients.
According to Figure~\ref{fig:spectral_radius_ol}, the spectral radii of the
plant systems identified with both methods share a similar trend. For low
regularization coefficients, the systems are correctly identified as unstable,
while for very high regularization coefficients, the plant systems are
incorrectly stabilized.
Figures~\ref{fig:spectral_radius_cl} and~\ref{fig:spectral_radius_ol} indicate
that a low regularization coefficient is appropriate for this system.

As a second point of comparison between the EDMD and the proposed method,
consider the three-fold cross-validation score associated with each
regularization coefficient.
A good cross-validation score should reach its maximum at an appropriate
hyperparameter value for the system. Given the spectral radius results in
Figures~\ref{fig:spectral_radius_cl} and~\ref{fig:spectral_radius_ol}, a good
scoring metric for the \textit{QUBE-Servo} system should have its peak at a low
regularization coefficient.
The scoring metric of choice for this system is the $R^2$ score, also called the
coefficient of determination~\cite{wright_1921_correlation}.
The $R^2$ score of a predicted state trajectory $\hat{x}_k$ relative to the true
state trajectory $x_k$ is
\begin{equation}
    R^2(x_k, \hat{x}_k)
    =
    1
    -
    \frac{\sum_{k=1}^{q}{(x_k - \hat{x}_k)}^2}{\sum_{k=1}^{q}{(x_k - \bar{x})}^2},
\end{equation}
where
$\bar{x} = \frac{1}{q} \sum_{k=1}^{q} x_k$ is the mean value of $x_k$.
For multidimensional predicted trajectories, the $R^2$ score of each state is
averaged to obtain a single score.
Perfect predictions receive an $R^2$ score of \num{1}, while predictions that
only capture the mean of the data receive an $R^2$ score of \num{0}. Worse
predictions can receive arbitrarily negative scores.

As before, Figure~\ref{fig:cross_validation_cl} shows the closed-loop $R^2$
scores of models identified with both EDMD and the proposed method, where the
plant model identified with EDMD is wrapped with the known controller. For any
regularization coefficient, the $R^2$ score achieved by the proposed method
remains bounded and positive. Thus, the closed-loop regularizer has the expected
effect of driving the predicted trajectory towards the mean as the
regularization coefficient grows.
However, for a large enough regularization coefficient, EDMD identifies an
unstable closed loop system and its $R^2$ score diverges. The best closed-loop
$R^2$ scores for both methods are attained with a very small regularization,
which also leads to the expected stability properties. The use of the plant's
open-loop predictions to score the models does not share this behaviour.
Since the plant system is inherently unstable, its predictions are highly
sensitive to small variations in parameters and initial conditions. Thus, even
the predictions of an accurate model can yield unbounded prediction errors.
As shown in Figure~\ref{fig:cross_validation_ol}, the only way to achieve a
finite plant $R^2$ score is to select an extremely large regularization
coefficient. For EDMD, this results in an unstable closed-loop system and an
asymptotically stable plant, which does not reflect the underlying dynamics of
the inverted pendulum system. While the proposed closed-loop methods produces an
asymptotically stable closed-loop system for all regularizers, the open-loop
plant it identifies still becomes asymptotically stable for a large enough
regularization coefficient.

Figure~\ref{fig:cross_validation} demonstrates two key conclusions. First, the
closed-loop prediction error should be used for assessing the accuracy of
identified models, as the best closed-loop scores correspond to Koopman models
with the expected stability properties.
Second, regularizing the closed-loop Koopman matrix is
preferable to regularizing only the plant's Koopman matrix, as it ensures the
closed-loop system will remain asymptotically stable for high regularization
coefficients.
Note that closed-loop scoring can be leveraged without the proposed method,
simply by identifying the plant using EDMD and wrapping the resulting model with
the known controller. However, this approach is susceptible to bias when
significant sensor noise is present.
Also, only the proposed closed-loop approach has access to the closed-loop
Koopman matrix for regularization.
Thus, if the controller is already known, it is preferable to use the proposed
closed-loop approach from the beginning, rather than leveraging controller
knowledge only at the end of the identification procedure.

\subsection{Comparison of optimized regularization coefficients}\label{sec:scoring}

\begin{figure}[htbp]
    \centering
    \begin{subfigure}[t]{3.5in}
        \centering
        \includegraphics[width=\linewidth]{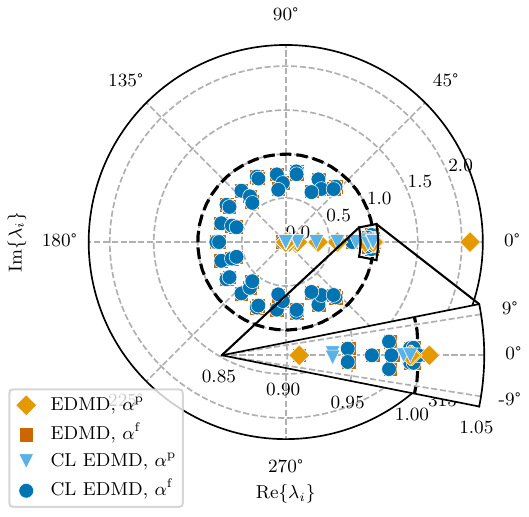}
        \caption{Eigenvalues of the CL systems identified with EDMD and CL
        EDMD.\@ Unlike EDMD, CL~EDMD correctly identifies a stable system for
        both regularization coefficients.}\label{fig:eigenvalues_cl}
    \end{subfigure}%
    \hfill
    \begin{subfigure}[t]{3.5in}
        \centering
        \includegraphics[width=\linewidth]{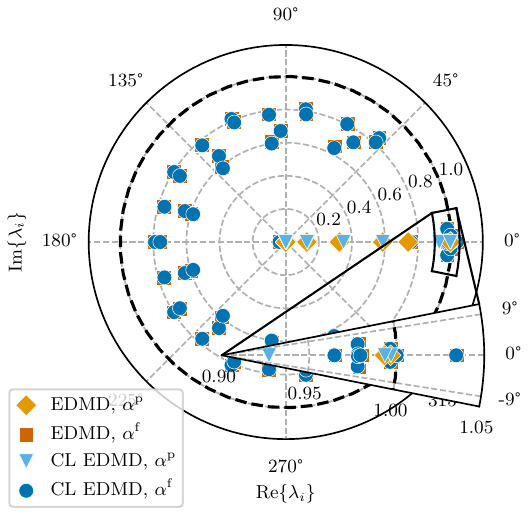}
        \caption{Eigenvalues of the plant systems identified with EDMD and CL
        EDMD.\@ For small regularization coefficients, both EDMD and CL~EDMD
        correctly identify unstable systems.}\label{fig:eigenvalues_ol}
    \end{subfigure}\vspace{2ex}
    \begin{subfigure}[t]{3.5in}
        \centering
        \includegraphics[width=\linewidth]{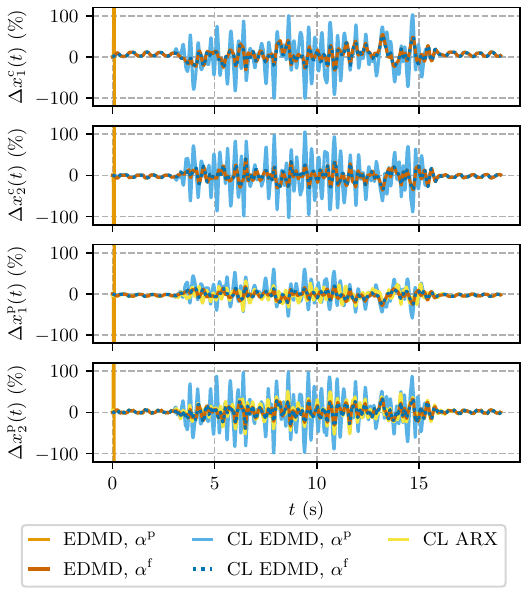}
        \caption{Prediction errors of the CL systems identified with EDMD and CL
        EDMD.\@ Unlike CL~EDMD, EDMD incorrectly identifies an unstable system
        when using a large regularization
        coefficient.}\label{fig:predictions_cl}
    \end{subfigure}%
    \hfill
    \begin{subfigure}[t]{3.5in}
        \centering
        \includegraphics[width=\linewidth]{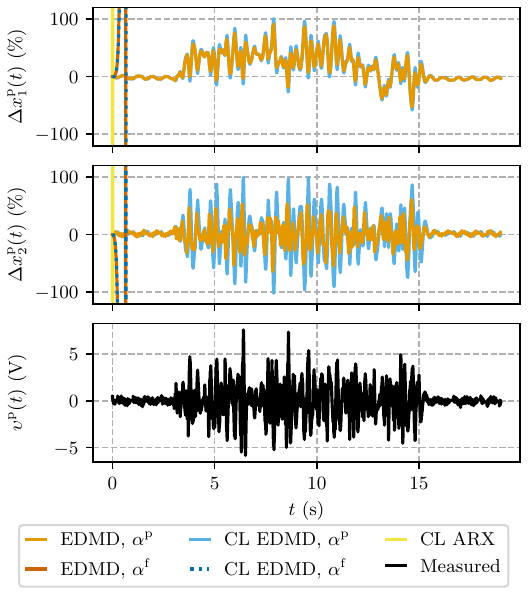}
        \caption{Prediction errors and controller output of the plant systems
        identified with EDMD and CL~EDMD.\@ Despite the diverging
        prediction errors, both methods correctly identify unstable plants for
        small regularization coefficients.}\label{fig:predictions_ol}
    \end{subfigure}
    \caption{Eigenvalues and prediction errors of closed-loop and plant
    systems.}\label{fig:predictions}
\end{figure}

Informed by the results of Section~\ref{sec:regularization}, four approaches
to identifying a Koopman model of the \textit{QUBE-Servo} are compared via
their eigenvalues and prediction errors.
First is the naive open-loop approach, where EDMD is used to identify the
plant's Koopman matrix, and the regularization coefficient is selected according
to the plant's $R^2$ score.
In this case, the optimal regularization coefficient is $\alpha^\mathrm{p} =
    10^{3}$.
Second is another open-loop approach, where EDMD is used to identify the
plant's Koopman matrix, but the plant is then wrapped with the known controller
and the regularization coefficient is selected according to the closed-loop
system's $R^2$ score.
In this case, the optimal regularization coefficient is $\alpha^\mathrm{f} =
    10^{-3}$.
Third is a closed-loop approach, where the closed-loop and plant's Koopman
matrices are simultaneously identified, but the plant's $R^2$ score is used to
select the regularization coefficient. In this case, the optimal regularization
coefficient is $\alpha^\mathrm{p} = 10^{3}$.
Last is the proposed closed-loop approach, where the closed-loop and plant's
Koopman matrices are simultaneously identified, and the closed-loop $R^2$ score
is used to select the regularization coefficient.
In this case, the optimal regularization coefficient is $\alpha^\mathrm{f} =
    10^{-3}$.
The prediction errors of these four Koopman models are also compared
with those of the closed-loop ARX (CL~ARX) approach described in
Section~\ref{sec:experimental_setup} without any regularization.

Consider the closed-loop and plant eigenvalues in
Figures~\ref{fig:eigenvalues_cl} and~\ref{fig:eigenvalues_ol}. The naive
approach, which uses the plant for regularization and scoring, identifies an
unstable closed-loop system with two eigenvalues clearly outside the unit
circle. Due to the large regularization coefficient, the plant system it
identifies is asymptotically stable, which is also inconsistent with the
underlying dynamics of the system.
The third approach, which uses the closed-loop system for regularization but the
plant for scoring, correctly identifies an asymptotically stable closed-loop
system, but incorrectly identifies an asymptotically stable plant.
The remaining two approaches, which use the closed-loop $R^2$ score to select
the regularization coefficient, identify essentially the same eigenvalues. Both
methods correctly identify asymptotically stable closed-loop systems and
unstable plant systems.
Since measurements from the \textit{QUBE-Servo} system are relatively noiseless,
it is expected that the two methods should agree. However, as is well-documented
in the system identification literature, the closed-loop method could reduce the
bias in the identified model in the presence of significant measurement
noise~\cite{forssell_1999_closedloop, hof_1998_closedloop},~\cite[\S11.1]{katayama_2005_subspace}.

Looking at the closed-loop and plant prediction errors of each of these
approaches in Figures~\ref{fig:predictions_cl} and~\ref{fig:predictions_ol}
respectively tells the same story as the eigenvalues. Both EDMD and the proposed
closed-loop method identify the same Koopman systems at low regularization
coefficients. For large regularization coefficients, EDMD incorrectly identifies
an unstable closed-loop system and an asymptotically stable plant.
However, the proposed method identifies an asymptotically stable closed-loop
system, regardless of the regularization coefficient.
The CL~ARX model correctly identifies an asymptotically stable
closed-loop system and an unstable open-loop system. In terms of closed-loop
prediction errors, the CL~ARX model is slightly worse than the Koopman models
that use closed-loop scoring, indicating that it may not be able to capture the
nonlinearity of the inverted pendulum system.
The $R^2$ scores and normalized root-mean-squared errors (NRMSE) of each model
on the test set are summarized in Table~\ref{tab:score}. To compute the NRMSE,
the root-mean-squared error of each state is normalized by the peak amplitude
of the true value of that state. The normalized values for each state are then
averaged to obtain a single number summarizing the error of the trajectory
prediction.

\begin{table}[ht]
    \centering
    \caption{$R^2$ score and NRMSE over \num{20} test
    episodes.}\label{tab:score}
    \begin{tabular}{lllcccc}
        \toprule
        Method &
        Regularization &
        Scoring &
        $R^2$ avg. &
        $R^2$ std. &
        NRMSE avg. &
        NRMSE std. \\
        \midrule
        ARX &
        --- &
        CL &
        \num{0.813} &
        \num{0.026} &
        \SI{11.7}{\percent} &
        \SI{1.3}{\percent} \\
        \midrule
        EDMD &
        Plant &
        Plant &
        $-\infty$ &
        --- &
        $\infty$ &
        --- \\
        \midrule
        EDMD &
        Plant &
        CL &
        \num{0.897} &
        \num{0.018} &
        \SI{8.9}{\percent} &
        \SI{1.1}{\percent} \\
        \midrule
        EDMD &
        CL &
        Plant &
        \num{0.322} &
        \num{0.061} &
        \SI{22.8}{\percent} &
        \SI{2.1}{\percent} \\
        \midrule
        EDMD &
        CL &
        CL &
        \num{0.893} &
        \num{0.017} &
        \SI{9.0}{\percent} &
        \SI{1.0}{\percent} \\
        \bottomrule
    \end{tabular}
\end{table}

Table~\ref{tab:score} shows that, among the Koopman models, the most
important factor for attaining a high prediction score is the use of the
closed-loop prediction error for scoring. For the \textit{QUBE-Servo} system,
the choice to use the closed-loop or plant Koopman matrix for regularization is
not critical, as long as a sufficiently small regularization coefficient is
chosen.
However, it must be noted that identifying the open-loop plant directly with
EDMD and wrapping the model with the known controller for scoring requires
knowledge of the controller and measurements of the controller output, just like
the proposed closed-loop approach.
If knowledge of the controller is already assumed, then using the proposed
closed-loop method is preferable, as it avoids the possibility of bias in the
models.
Furthermore, in the proposed closed-loop Koopman operator approximation approach
both $\mbf{U}^\mathrm{f}$ and $\mbf{U}^\mathrm{p}$ are available as optimization
variables. This additional flexibility allows additional knowledge of the
closed-loop system and plant to be incorporated into the optimization problem.
For example, the closed-loop system could be constrained to be asymptotically
stable~\cite{dahdah_system_2022}. Alternatively, known properties of the
closed-loop system or plant could be incorporated by constraining the poles to a
particular region~\cite{chilali_1999_robust}.
A distinct advantage of CL~EDMD over CL~ARX is that it is
constraint-based, meaning that it does not require transfer matrix
multiplication or pseudoinversion to obtain an estimate of the plant system.
Due to the nature of the constraints, the identified plant system will always
be of a lower dimension than the closed-loop system.
The advantages and disadvantages of each approach are summarized in
Table~\ref{tab:pro_con}.

\begin{table}[ht]
    \centering
    \caption{Comparison of system identification methods.}\label{tab:pro_con}
    \begin{tabular}{lllccccc}
        \toprule
        Method &
        Regularization &
        Scoring &
        Bounded score? &
        CL regularizer? &
        Avoids bias? &
        Nonlinear? \\
        \midrule
        ARX &
        --- & CL &
        \cmark{} &
        --- &
        \cmark{} &
        \xmark{} \\
        \midrule
        EDMD &
        Plant & Plant &
        \xmark{} &
        \xmark{} &
        \xmark{} &
        \cmark{} \\
        \midrule
        EDMD &
        Plant & CL &
        \cmark{} &
        \xmark{} &
        \xmark{} &
        \cmark{} \\
        \midrule
        EDMD &
        CL & Plant &
        \xmark{} &
        \cmark{} &
        \cmark{} &
        \cmark{} \\
        \midrule
        EDMD &
        CL & CL &
        \cmark{} &
        \cmark{} &
        \cmark{} &
        \cmark{} \\
        \bottomrule
    \end{tabular}
\end{table}

\section{Conclusion}\label{sec:conclusion}

When identifying closed-loop systems, it is often not possible to neglect the
effects of the control loop. This holds true for Koopman operator approximation
as well as linear system identification. In this paper, a closed-loop Koopman
operator identification method is presented, where the closed-loop system and
plant system are identified simultaneously using knowledge of the controller.
The advantages of this method is demonstrated in simulation using a
Duffing oscillator with coloured measurement noise and experimentally using an
unstable rotary inverted pendulum system. In particular, the proposed
closed-loop method identifies Koopman models that align with the prior
knowledge that the closed-loop system is asymptotically stable, while the plant
is unstable. The closed-loop method also naturally allows the closed-loop
prediction error to be used as a goodness-of-fit metric, which is crucial for
selecting lifting functions and an appropriate regularization coefficient.
Furthermore, by identifying the plant system using constraints, the
proposed closed-loop approach avoids inflating the dimension of the plant
system.

A limitation of the proposed approach is that including the controller
state in the lifted state of the system increases its dimension, especially if
complex controllers are used. While high-order controllers are generally
undesirable in system identification applications, the assumption that the
controller is known exactly does not always hold. Unmodelled effects like
actuator saturation may limit a user's ability to correctly calculate the
controller state, which could affect the accuracy of the method.

The proposed closed-loop Koopman operator approximation method allows both the
closed-loop and plant Koopman matrices to be used in regularizers or
constraints. Using Extended DMD, only the plant's Koopman matrix can be used in
a regularizer, leading to potentially unstable closed-loop systems for large
regularization coefficients. The additional flexibility provided by the
proposed closed-loop Koopman operator approximation method will be explored in
future work, as the ability to incorporate known information about the
closed-loop system or plant into the regression problem could prove useful in
identifying more accurate or useful Koopman models. Extensions to address the
use of nonlinear controllers, or to address the situation where the controller
is also unknown could also prove valuable in broadening the applicability of
the proposed closed-loop Koopman operator approximation method.

\section*{Acknowledgements}\label{sec:acknowledgements}

This work is supported by the Natural Sciences and Engineering Research Council
of Canada (NSERC) Discovery Grants program, the \textit{Institut de valorisation
    des donn{\'e}es} (IVADO), the Canadian Institute for Advanced Research (CIFAR),
and the \textit{Centre de recherches math{\'e}matiques} (CRM), as well as by
Mecademic through the Mitacs Accelerate program.
The authors thank Quanser for the use of the \textit{QUBE-Servo} rotary
inverted pendulum system.

\section*{Data availability statement}\label{sec:data_availability}

The data that support the findings of this study are openly available.

\appendix
\section{Recovery of the plant using least-squares}\label{sec:lstsq}

\begin{figure}[htbp]
    \centering
    \begin{subfigure}[t]{3.5in}
        \centering
        \includegraphics[width=\linewidth]{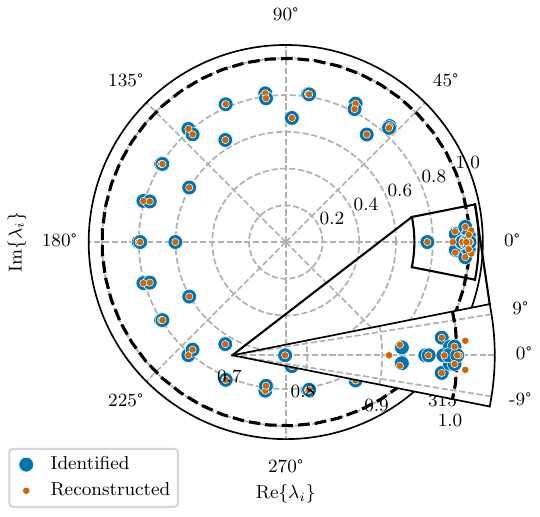}
        \caption{CL eigenvalues identified with the least-squares version of CL
        EDMD, before and after the plant is extracted and the controller is
        re-connected. When the CL system is reconstructed from the plant, the
        eigenvalues move, and some of them leave the unit circle, destabilizing
        the system.}\label{fig:controller_rewrap_eig_lstsq}
    \end{subfigure}%
    \hfill
    \begin{subfigure}[t]{3.5in}
        \centering
        \includegraphics[width=\linewidth]{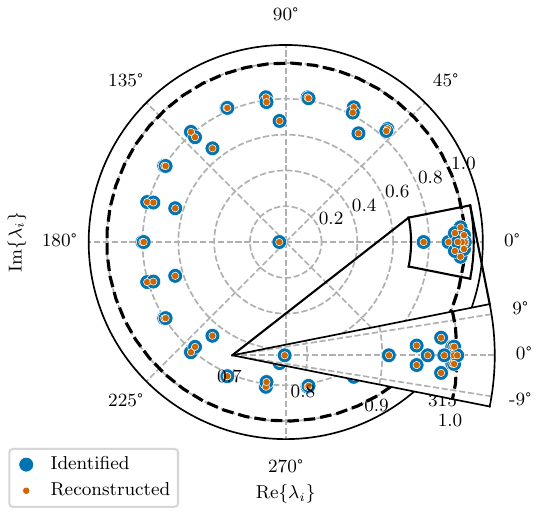}
        \caption{CL eigenvalues identified with the constrained version of CL
        EDMD, before and after the plant is extracted and the controller is
        re-connected. When the CL system is reconstructed from the plant, the
        eigenvalues do not change.}\label{fig:controller_rewrap_eig_const}
    \end{subfigure}\vspace{2ex}
    \begin{subfigure}[t]{3.5in}
        \centering
        \includegraphics[width=\linewidth]{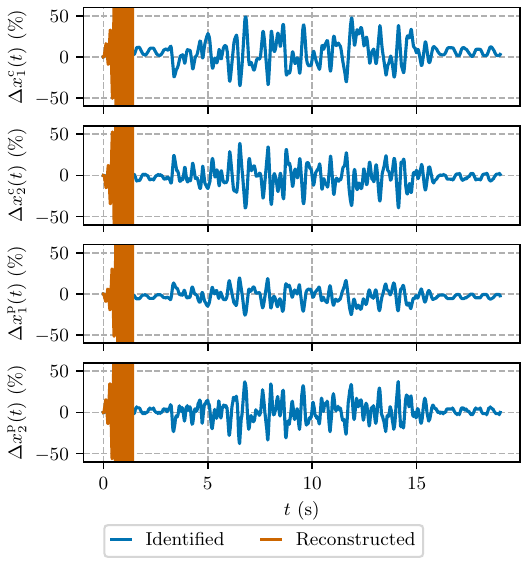}
        \caption{CL prediction errors of models identified with
        the least-squares version of CL~EDMD, before and after the plant is
        extracted and the controller is re-connected. Reconstructing the CL
        system destabilizes it.}\label{fig:controller_rewrap_pred_lstsq}
    \end{subfigure}%
    \hfill
    \begin{subfigure}[t]{3.5in}
        \centering
        \includegraphics[width=\linewidth]{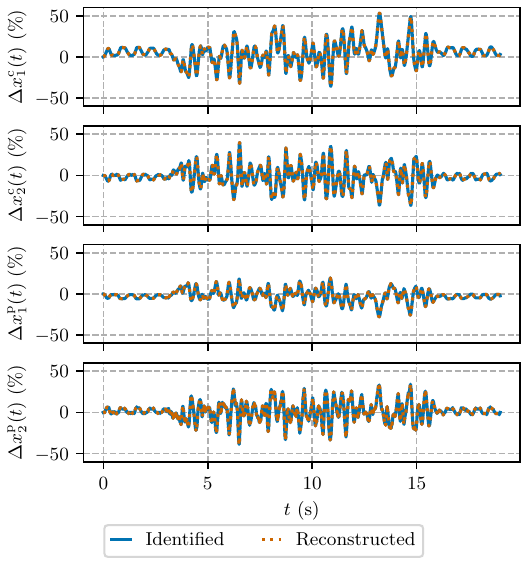}
        \caption{CL prediction errors of models identified with the constrained
        version of CL~EDMD, before and after the plant is extracted and the
        controller is re-connected. Reconstructing the CL system does not change
        its predictions.}\label{fig:controller_rewrap_pred_const}
    \end{subfigure}
    \caption{CL eigenvalues and prediction errors of models identified
    with the least-squares and constrained versions of CL
    EDMD.}\label{fig:controller_rewrap_pred}
\end{figure}

Using least-squares to extract a plant model from an identified closed-loop
system seems to provide a simpler alternative to the constraint-based
formulation in~\eqref{eq:opt_abstract_cost}--\eqref{eq:opt_abstract_Ap}.
However, using~\eqref{eq:lstsq_Bp} and~\eqref{eq:constr_Ap} to obtain a Koopman
model of the plant does not respect the feedback structure of the system. In
fact, extracting the plant from the closed-loop system and re-wrapping it with
the same controller does not result in the same system.
Figure~\ref{fig:controller_rewrap_eig_lstsq} shows how the closed-loop
eigenvalues change when the plant is extracted and re-wrapped with a controller.
In this case, the procedure actually destabilizes the system. When using the
constraint-based approach, as shown in
Figure~\ref{fig:controller_rewrap_eig_const}, extracting the plant and closing
the loop again with the same controller does not move the eigenvalues.
Figures~\ref{fig:controller_rewrap_pred_lstsq}
and~\ref{fig:controller_rewrap_pred_const}
show the prediction errors of each
system before and after removing and re-adding the controller. As expected, the
predicted trajectories of the destabilized system diverge quickly.
The ability to extract the plant from the closed-loop system while respecting
the feedback structure of the system is particularly important for control
design tasks, where, presumably, the plant is being identified with the end goal
of designing an improved controller.

\bibliographystyle{unsrt}
\bibliography{paper-v3}

\end{document}